\documentclass[english,notitlepage,nofootinbib]{revtex4-1}
\pdfoutput=1
\usepackage[T1]{fontenc}
\usepackage[latin9]{inputenc}
\usepackage{color}
\usepackage{graphicx}
\usepackage{esint}
\usepackage{afterpage}
\usepackage{rotating}
\usepackage[percent]{overpic}

\makeatletter

\@ifundefined{textcolor}{}
{%
 \definecolor{BLACK}{gray}{0}
 \definecolor{WHITE}{gray}{1}
 \definecolor{RED}{rgb}{1,0,0}
 \definecolor{GREEN}{rgb}{0,1,0}
 \definecolor{BLUE}{rgb}{0,0,1}
 \definecolor{CYAN}{cmyk}{1,0,0,0}
 \definecolor{MAGENTA}{cmyk}{0,1,0,0}
 \definecolor{YELLOW}{cmyk}{0,0,1,0}
}

\makeatother

\usepackage{babel}
\begin{document}

\title{Sterile neutrinos in the light of IceCube}

\date{\today}

\author{Manfred Lindner$^a$, Werner Rodejohann$^a$ and Xun-Jie Xu$^{a,b}$, }

\affiliation{$^a$Max-Planck-Institut f\"{u}r Kernphysik, Postfach 103980, D-69029
Heidelberg, Germany\\
$^b$Institute of Modern Physics and Center for High Energy Physics,
Tsinghua University, Beijing 100084, China}

\begin{abstract}
\noindent
We determine  constraints on parameters of a single 
eV-scale light neutrino using IceCube-59 data. Particular 
emphasis is put on the question whether such an analysis can rule out 
sterile neutrino hints. While important 
complementary information is provided, the different dependence 
on the various sterile neutrino mixing angles makes it currently not 
possible to fully exclude short baseline appearance results or sterile neutrinos in general.

\end{abstract}
\maketitle

\section{Introduction}

Though the standard 3-neutrino mixing paradigm has been well established \cite{Agashe:2014kda}, 
there are still several short-baseline anomalies, most notably 
in LSND \cite{Aguilar:2001ty}, MiniBooNE \cite{Aguilar-Arevalo:2013pmq} and 
reactor neutrino flux measurements \cite{Mueller:2011nm,Huber:2011wv}. 
These anomalies could be explained by introducing light sterile
neutrinos with the mass-squared difference $\Delta m^{2}$ at $0.1\sim1\,{\rm eV}^{2}$
scale and mixing matrix elements around 0.1. If such light sterile neutrinos would indeed 
exist, the theoretical implications would be profound. Therefore 
several experiments are running or in construction in order to confirm or refuse 
the existence of light sterile neutrinos. See the recent 
reviews \cite{Abazajian:2012ys,Gariazzo:2015rra} for an overview of the hints, consequences 
and tests of light sterile neutrinos. 

In this work we focus on effects of eV-scale sterile neutrinos in atmospheric 
neutrino oscillations at high energies as measured in the IceCube experiment. 
We use here an IceCube-59 data set from Ref.\ \cite{Aartsen:2013eka}, where a 
search for diffuse astrophysical neutrinos was performed. 
The fact that sterile neutrinos would 
have an impact in IceCube is easily understood by noting that a mass-squared difference 
of order eV$^2$ corresponds to maximal oscillations at energies 
$E_{\nu}\sim10^{3}$ GeV and a baseline around 
Earth radius $R_{\oplus}\simeq6.4\times10^{3}$ km. Indeed, atmospheric
neutrinos observed in IceCube have energies ranging from $10^{2}$ GeV
to $10^{6}$ GeV, peaked at about $10^{3}$ GeV. Several papers have in the past analyzed the 
effect of light sterile neutrinos at high energies as a potential test of the 
sterile neutrino hypothesis \cite{Nunokawa:2003ep,Choubey:2007ji,Razzaque:2011ab,Barger:2011rc,Esmaili:2012nz,Esmaili:2013cja,Esmaili:2013vza,Arguelles:2012cf}.  

We will perform here 
a $\chi^{2}$-analysis on the IceCube-59 data within a $3+1$ scheme to see how 
significant the constraint on sterile neutrinos is\footnote{
Here the $3+1$ scheme refers to the case that there are 3 active
neutrinos and 1 heavier sterile neutrino. In principle other cases
such as 1 sterile neutrino $+$ 3 heavier active neutrinos, 3 active
neutrinos $+$ 2 heavier sterile neutrinos, are also possible. However, 
since in these cases the total neutrino mass $\Sigma_{i} m_{i}$ is much 
larger, they are disfavored by cosmological constraints.}. 
We are particularly interested in the 
interplay of this constraint with the results of short baseline appearance and other 
experiments. We work in the minimal framework of only one sterile neutrino with a unitary 
$4\times4$ matrix and no additional interactions. Even in this simple situation 
the dependence on the various lepton mixing angles is different in each experiment, and  
in addition, muon neutrino disappearance in IceCube depends on the weakly constrained 
angle $\theta_{34}$, that should not be set to zero in analyses. 
For scenarios with more than one sterile state the larger number of mixing angles will complicate further the direct comparison of oscillation probabilities in IceCube and other experiments. 
Moreover, there is dependence on the sterile neutrino global fit results to which one compares the IceCube sensitivity.  
All in all, a full exclusion of short baseline appearance and/or other 
results is currently not possible, though of course important complementary 
constraints on sterile neutrinos are provided by IceCube data. \\

The paper is build up as follows: in Section \ref{sec:nu} we discuss 
the procedure to obtain the oscillation probability of high energy neutrinos 
including matter effects, before discussing 
event numbers in IceCube in Section \ref{sec:atm}. The numerical analysis of IceCube 
data is performed in Section \ref{sec:chi}, where we also discuss the 
comparison of the parameters 
crucial for IceCube with the ones for short baseline appearance and other experiments.

\section{Atmospheric neutrino disappearance induced by sterile neutrinos}
\label{sec:nu}
Neutrino oscillation can be described by the following Schr\"{o}dinger
equation in flavor space, 
\begin{equation}
i\frac{d}{dL}|\nu(L)\rangle=H|\nu(L)\rangle,\label{eq:1228}
\end{equation}
where $H$ is the effective Hamiltonian and $|\nu(L)\rangle$ denotes
the flavor state of the  neutrino at a distance of $L$ from
the source. For the standard 3+1 neutrino framework with 3 active neutrinos 
$(\nu_{e},\nu_{\mu},\nu_{\tau})$ 
plus one sterile neutrino $\nu_{s}$, including matter effects, the effective Hamiltonian $H$ has the following matrix form
\begin{equation}
H=\frac{1}{2E}U\left(\begin{array}{cccc}
m_{1}^{2}\\
 & m_{2}^{2}\\
 &  & m_{3}^{2}\\
 &  &  & m_{4}^{2}
\end{array}\right)U^{\dagger}+\sqrt{2}G_{F}N_{e}\left(\begin{array}{cccc}
1\\
 & 0\\
 &  & 0\\
 &  &  & \kappa
\end{array}\right).\label{eq:1227-1}
\end{equation}
Here $E$ is the neutrino energy, $m_{i}$  and $U$ are the neutrino
masses and mixing matrix; $G_{F}$ is the Fermi constant, $N_{e}$
is the electron number density of matter and $\kappa$ is a ratio defined
as 
\begin{equation}
\kappa\equiv\frac{N_{n}}{2N_{e}}\,,\label{eq:0906}
\end{equation}
where $N_{n}$ is the neutron number density. For anti-neutrinos,
we need to replace $U\rightarrow U^{*}$ and $G_{F}N_{e}\rightarrow-G_{F}N_{e}$
in Eq.\ (\ref{eq:1227-1}). 

For constant matter density, $H$ does not vary with $L$ so the
solution is simply $|\nu(L)\rangle=e^{-iHL}|\nu(0)\rangle$. But
the matter density of Earth \cite{Dziewonski:1981xy} varies significantly from $1\,\rm g/{\rm cm}^{3}$ (at the
surface) to $13\,\rm g/{\rm cm}^{3}$ (at the inner core), so that to obtain accurate
results one has to either numerically solve the full differential 
equation Eq.\ (\ref{eq:1228}), or divide the full neutrino path into many 
segments with approximated constant densities so that 
\begin{equation}
|\nu(L)\rangle=\prod_{i}e^{-iH_{i}L_{i}}|\nu(0)\rangle\label{eq:0906-1}\,.
\end{equation}
Here $H_{i}$ is the Hamiltonian in the $i$-th segment and $L_{i}$
is the corresponding baseline. Actually this is the main method
used to compute oscillation probabilities in the GLoBES package 
\cite{Huber:2007ji} (see also \cite{Wallraff:2014qka}). In
this work, we will adopt the same method, i.e.\ Eq.\ (\ref{eq:0906-1}), 
to compute probabilities. Defining the $S$-matrix as 
\begin{equation}
S_{\alpha\beta}\equiv \left(\prod_{i}e^{-iH_{i}L_{i}}\right)_{\alpha\beta}\,,\label{eq:0907}
\end{equation}
where $\alpha$, $\beta$ are flavor indices ($e$, $\mu$, $\tau$
and $s$), the survival probability of $\nu_{\mu}$ is given
by 
\begin{equation}
P_{\nu_{\mu}\rightarrow\nu_{\mu}}=|S_{\mu\mu}|^{2}\,.\label{eq:0907-1}
\end{equation}
For the energy range of IceCube, $\nu_{\mu}$ and $\overline{\nu}_{\mu}$
dominate the atmospheric neutrino flux while $\nu_{e}$ and $\overline{\nu}_{e}$
are negligible\footnote{
Compared to the $\nu_{\mu}$ ($\overline{\nu}_{\mu}$)
flux, the $\nu_{e}$ ($\overline{\nu}_{e}$) flux is suppressed by
a factor of about 10 \cite{Honda:2006qj,Fedynitch:2012fs}. Moreover, clean muon-neutrino samples can be obtained by 
observing the corresponding muon track.
}; thus only $P_{\nu_{\mu}\rightarrow\nu_{\mu}}$ and 
$P_{\overline{\nu}_{\mu}\rightarrow\overline{\nu}_{\mu}}$
will be used in this work. The $4\times4$ mixing matrix is 
\[
U=R^{34}R^{24}R^{14}R^{23}R^{13}R^{12}\,,
\]
where $R^{ij}$ is a $4\times4$ matrix whose $(i,i)$, $(i,j)$,
$(j,i)$ and $(j,j)$ elements are $\cos\theta_{ij}$, $\sin\theta_{ij}e^{-i\delta_{ij}}$,
$-\sin\theta_{ij}e^{i\delta_{ij}}$ and $\cos\theta_{ij}$, respectively.
The other elements are the same as for a $4\times4$ identity matrix.  Actually
there are only three independent CP-violating phases relevant for neutrino oscillations 
and one possible convention is to set $\delta_{34}=\delta_{23}=\delta_{12}=0$\footnote{Keeping 
all six phases is useful because then the three Majorana phases are automatically 
taken care of as well \cite{Rodejohann:2011vc}.}.   
Note that in the limit of vanishing atmospheric and solar mass-squared differences 
there is no CP effect, and since these two mass-squared differences are much smaller than the one corresponding 
to sterile neutrinos, the effect of CP phases is suppressed.
Therefore in our analysis, we will neglect them. 
If all CP-violating phases are zero, the last column of $U$, denoted
as $u_{4}$, which we will need later in this analysis, has the following form: 
\begin{equation}
u_{4}=(\sin\theta_{14},\thinspace\cos\theta_{14}\sin\theta_{24},\thinspace\cos\theta_{14}\cos\theta_{24}\sin\theta_{34},\thinspace\cos\theta_{14}\cos\theta_{24}\cos\theta_{34})^{T}\,.\label{eq:0907-2}
\end{equation}
Let us now discuss matter effects. For eV-scale neutrinos with TeV-scale energies, 
the first and second terms in 
Eq.\ (\ref{eq:1227-1}) are comparable, about $m_{4}^{2}/2E\sim\sqrt{2}G_{F}N_{e}\sim10^{-13}$ eV. 
Therefore one expects that matter effects have a significant
influence on the probability. As usual for matter effects, a resonance can appear for certain 
values of energy and baseline. For the case under study, this can only happen for 
the $\overline{\nu}_{\mu}\rightarrow\overline{\nu}_{\mu}$ channel, as has been recently studied 
for instance in Ref.\ \cite{Esmaili:2013vza}. As a result of the resonance, the survival 
probability can become zero; 
anti-muon-neutrinos would completely disappear, even if the active-sterile
mixing angles are small.  Hence matter effects are crucial for the sensitivity of IceCube 
to light sterile neutrinos. 

However, a less noticed point is that though the matter effect contribution
to the effective Hamiltonian is large, in some case, $\overline{\nu}_{\mu}$
and $\nu_{\mu}$ may oscillate as if they are in vacuum. The condition
for this case is that $U_{s4}$ is zero or small, or equivalently that $\theta_{34}$ is $\pi/2$ or large.  We first show this analytically with the single mass-squared-difference approximation
in constant density matter and then numerically verify it by taking
into account all mass-squared differences and also density variation. 
Considering that $\Delta m_{41}^{2}\gg\Delta m_{31}^{2},\thinspace\Delta m_{21}^{2}$
we can neglect the effect of $\Delta m_{31}^{2},\thinspace\Delta m_{21}^{2}$
and set them to zero. Defining 
\begin{equation}
\Delta m_{41}^{2}\equiv\Delta m^{2}\mbox{ and }
A\equiv2\sqrt{2}G_{F}N_{e}E/\Delta m^{2}\,,\label{eq:0908-1}
\end{equation}
we can write $H$ as follows:  
\begin{equation}
H=\frac{m_{1}^{2}}{2E}+\frac{\Delta m^{2}}{2E}M\,.\label{eq:0908-2}
\end{equation}
Here $M$ is a dimensionless matrix, 
\begin{equation}
M\equiv u_{4}.u_{4}^{T}+A\left(\begin{array}{cccc}
1\\
 & 0\\
 &  & 0\\
 &  &  & \kappa
\end{array}\right),\label{eq:0908-3}
\end{equation}
with $u_{4}$ being the last column of $U$, see Eq.\ (\ref{eq:0907-2}). Note that $u_{4}^{T}$
above would be $u_{4}^{\dagger}$ if $u_{4}$ was complex. But since
rephasing $M$ by $M\rightarrow QMQ^{\dagger}$ with $Q={\rm diag}(e^{i\alpha_{1}},e^{i\alpha_{2}},e^{i\alpha_{3}},e^{i\alpha_{4}})$
does not have any physical effect, we can always make $u_{4}$ real 
by such a transformation. This also implies that all CP-violating
phases in the mixing matrix are negligible if $\Delta m_{31}^{2},\thinspace\Delta m_{21}^{2}$
are negligible, as it should be. 
The $S$-matrix, assuming constant density, is 
\begin{equation}
S=e^{-iHL}=e^{-i\frac{m_{1}^{2}L}{2E}}e^{-itM},\label{eq:0108-1}
\end{equation}
where $t=\frac{\Delta m^{2}L}{2E}$. The overall phase $e^{-i\frac{m_{1}^{2}L}{2E}}$
does not affect the probability so it can be ignored. 
Considering the case $U_{s4}=0$, we can write $u_{4}$ as
\begin{equation}
u_{4}=(e_{4},\mu_{4},\tau_{4},0)^{T}\,,\label{eq:0908-4}
\end{equation}
where $e_{4}$, $\mu_{4}$ and $\tau_{4}$ are short for $U_{e4}$,
$U_{\mu4}$ and $U_{\tau4}$. Now $M$ is a block-diagonal $4\times4$ matrix:  
\begin{equation}
e^{-itM}=\left(\begin{array}{cc}
e^{-itM_{3}} & 0\\
0 & e^{-itA\kappa}
\end{array}\right),\label{eq:0908-5}
\end{equation}
where $M_{3}$ is a $3\times3$ matrix, 
\begin{equation}
M_{3}\equiv\left(\begin{array}{c}
e_{4}\\
\mu_{4}\\
\tau_{4}
\end{array}\right)\left(\begin{array}{ccc}
e_{4} & \mu_{4} & \tau_{4}\end{array}\right)+A\left(\begin{array}{ccc}
1\\
 & 0\\
 &  & 0
\end{array}\right).\label{eq:0908-6}
\end{equation}
The $\nu_{\mu}$ survival probability $P_{\nu_{\mu}\rightarrow\nu_{\mu}}=|S_{\mu\mu}|^{2}$
and $S_{\mu\mu}=(e^{-itM_{3}})_{\mu\mu}$ can be computed as follows:  
According to the Cayley-Hamilton theorem, $e^{-itM_{3}}$ can be written
as
\begin{equation}
e^{-itM_{3}}=s_{0}I+s_{1}M_{3}+s_{2}M_{3}^{2}\,.\label{eq:0101}
\end{equation}
Here the coefficients $s_{0}$, $s_{1}$ and $s_{2}$ are functions
of the three eigenvalues (see e.g. \cite{Xu:2015kma}), 
\begin{eqnarray}
s_{0} & = & \frac{-1}{\Delta_{\lambda}}[e^{-it\lambda_{3}}\lambda_{1}\lambda_{2}(\lambda_{1}-\lambda_{2})+e^{-it\lambda_{1}}\lambda_{2}\lambda_{3}(\lambda_{2}-\lambda_{3})+e^{-it\lambda_{2}}\lambda_{1}\lambda_{3}(\lambda_{3}-\lambda_{1})]\,,\label{eq:0101-1}\\
s_{1} & = & \frac{1}{\Delta_{\lambda}}[e^{-it\lambda_{3}}(\lambda_{1}^{2}-\lambda_{2}^{2})+e^{-it\lambda_{1}}(\lambda_{2}^{2}-\lambda_{3}^{2})+e^{-it\lambda_{2}}(\lambda_{3}^{2}-\lambda_{1}^{2})]\,,\label{eq:0101-2}\\
s_{2} & = & \frac{-1}{\Delta_{\lambda}}[e^{-it\lambda_{3}}(\lambda_{1}-\lambda_{2})+e^{-it\lambda_{1}}(\lambda_{2}-\lambda_{3})+e^{-it\lambda_{2}}(\lambda_{3}-\lambda_{1})]\,,\label{eq:0101-3}
\end{eqnarray}
where $(\lambda_{1},\lambda_{2},\lambda_{3})$ are the three eigenvalues of $M_{3}$
and $\Delta_{\lambda}\equiv(\lambda_{1}-\lambda_{2})(\lambda_{2}-\lambda_{3})(\lambda_{3}-\lambda_{1})$.
The eigenvalues of $M_{3}$ are
\begin{equation}
\lambda_{1,2,3}=0,\thinspace\frac{1}{2}\left(1+A\pm C\right),\label{eq:0909}
\end{equation}
where
\begin{equation}
C=\sqrt{4Ae_{4}^{2}+(A-1)^{2}}.\label{eq:0909-1}
\end{equation}
Combine all these result we get 
\begin{equation}
S_{\mu\mu}=\frac{\mu_{4}^{2}e^{-\frac{1}{2}it(A+C+1)}\left[-e^{Cit}\left(A+C+2e_{4}^{2}-1\right)+A-C+2e_{4}^{2}-1\right]+2C(e_{4}^{2}+\mu_{4}^{2}-1)}{2(e_{4}^{2}-1)C},
\label{eq:0908-8}
\end{equation}
The expansion of Eq.\ (\ref{eq:0908-8}) in small $e_{4}^{2} = |U_{e4}|^2$ gives
\begin{equation}
S_{\mu\mu}=1-\left(1-e^{-it}\right)\mu_{4}^{2}+\frac{e^{-iAt}-(A-1)^{2}-Ae^{-it}(2+it-A(1+it))}{(A-1)^{2}}\mu_{4}^{2}e_{4}^{2}+ {\cal O}(e_{4}^{4})\,.\label{eq:0908-7}
\end{equation}
Note that for a typical matter density of $\rho=6.5\,\rm g/{\rm cm}^{3}$ and 
$\Delta m^{2}=1\,{\rm eV}^{2}$, $A$ is about $0.5E/{\rm TeV}$. So in the energy range from 
$(10^{2}\sim10^{4})$ GeV, it is quite typical for $A$ to be $1$ (or close to 1) and 
the denominator in Eq.\ (\ref{eq:0908-7}) would be 0 (or close to 0). However, in this
case Eq.\ (\ref{eq:0908-7}) is still valid and accurate since the coefficient
before $\mu_{4}^{2}e_{4}^{2}$ will not blow up when $A\rightarrow1$, 
as one can check directly. Actually the singularity here corresponds
to a branch cut singularity and as it has been proved in Ref.\ \cite{Xu:2015kma},
all branch cut singularities should cancel out in the $S$-matrix. 
This is the deeper reason of the good behavior of Eq.\ (\ref{eq:0908-7})
at $A\rightarrow1$. Therefore, the  coefficient before
$\mu_{4}^{2}e_{4}^{2}$ can be regarded as an ${\cal O}(1)$ number that 
varies with $t$ (i.e.\ with $L/E$). 

If we take the vacuum limit $A=0$ we obtain 
\begin{equation}
S_{\mu\mu}^{{\rm vac}}=1-\left(1-e^{-it}\right)\mu_{4}^{2}\,,\label{eq:0908-10}
\end{equation}
and therefore Eq.\ (\ref{eq:0908-7}) can be written as
\begin{equation}
S_{\mu\mu}=S_{\mu\mu}^{{\rm vac}}+{\cal O}(1)\mu_{4}^{2}e_{4}^{2}+{\cal O}(e_{4}^{4})\,.\label{eq:0908-11}
\end{equation}
Eq.\ (\ref{eq:0908-11}) has an important implication. Since $e_{4}^{2}=|U_{e4}|^{2}$
has been constrained by reactor neutrino experiments to be small, 
typically less than $s_{13}^{2}$, the difference 
between $S_{\mu\mu}$ and $S_{\mu\mu}^{{\rm vac}}$ is small. 
We thus reach the conclusion that if $U_{s4}=0$, $\nu_{\mu}$($\overline{\nu}_{\mu}$)
will oscillate as if they would propagate in vacuum. Recalling that the resonance of 
the matter effect is crucial 
for the constraints on sterile neutrinos, it implies that the value of 
$U_{s4}$ or $\theta_{34}$ is important for the constraints, and eventually on the ability of 
IceCube data to rule out sterile neutrino hints. \\

\begin{figure}[t]
\centering
\begin{overpic}[width=16cm]{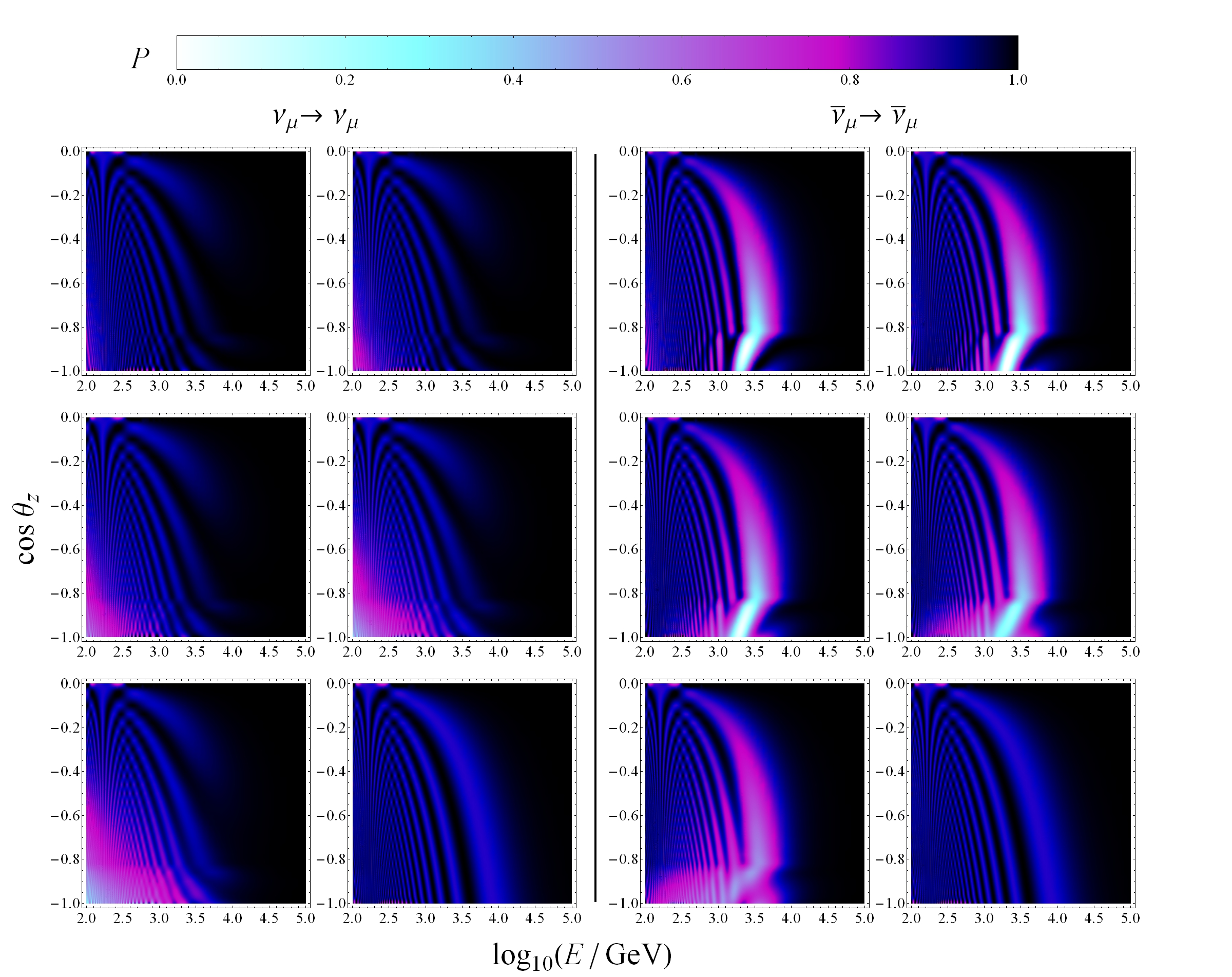}
\put (16,65) {\color{white} $\theta_{34}=0^{\circ}$}
\put (37,65) {\color{white} $\theta_{34}=10^{\circ}$}
\put (16,43.4) {\color{white} $\theta_{34}=20^{\circ}$}
\put (37,43.4) {\color{white} $\theta_{34}=30^{\circ}$}
\put (16,21.8) {\color{white} $\theta_{34}=45^{\circ}$}
\put (37,21.8) {\color{white} $\theta_{34}=90^{\circ}$}

\put (61.4,65) {\color{white} $\theta_{34}=0^{\circ}$}
\put (82.4,65) {\color{white} $\theta_{34}=10^{\circ}$}
\put (61.4,43.4) {\color{white} $\theta_{34}=20^{\circ}$}
\put (82.4,43.4) {\color{white} $\theta_{34}=30^{\circ}$}
\put (61.4,21.8) {\color{white} $\theta_{34}=45^{\circ}$}
\put (82.4,21.8) {\color{white} $\theta_{34}=90^{\circ}$}
\end{overpic}

\caption{\label{fig:The-survival-probablity}
The survival probabilities $P_{\nu_{\mu}\rightarrow\nu_{\mu}}$ 
(left panel) and $P_{\overline{\nu}_{\mu}\rightarrow\overline{\nu}_{\mu}}$
(right panel) in the 3(active)+1(sterile) scheme.  
We assume  
$\theta_{34}=(0^{\circ}, 10^{\circ}, 20^{\circ}, 30^{\circ}, 45^{\circ}, 90^{\circ})$, increased from left to right,
 top to bottom. Other parameters are $\Delta m^{2}=0.8\,{\rm eV}^{2}$,
$\theta_{24}=10^{\circ}$ and $\theta_{14}=0^{\circ}$. }

\end{figure}
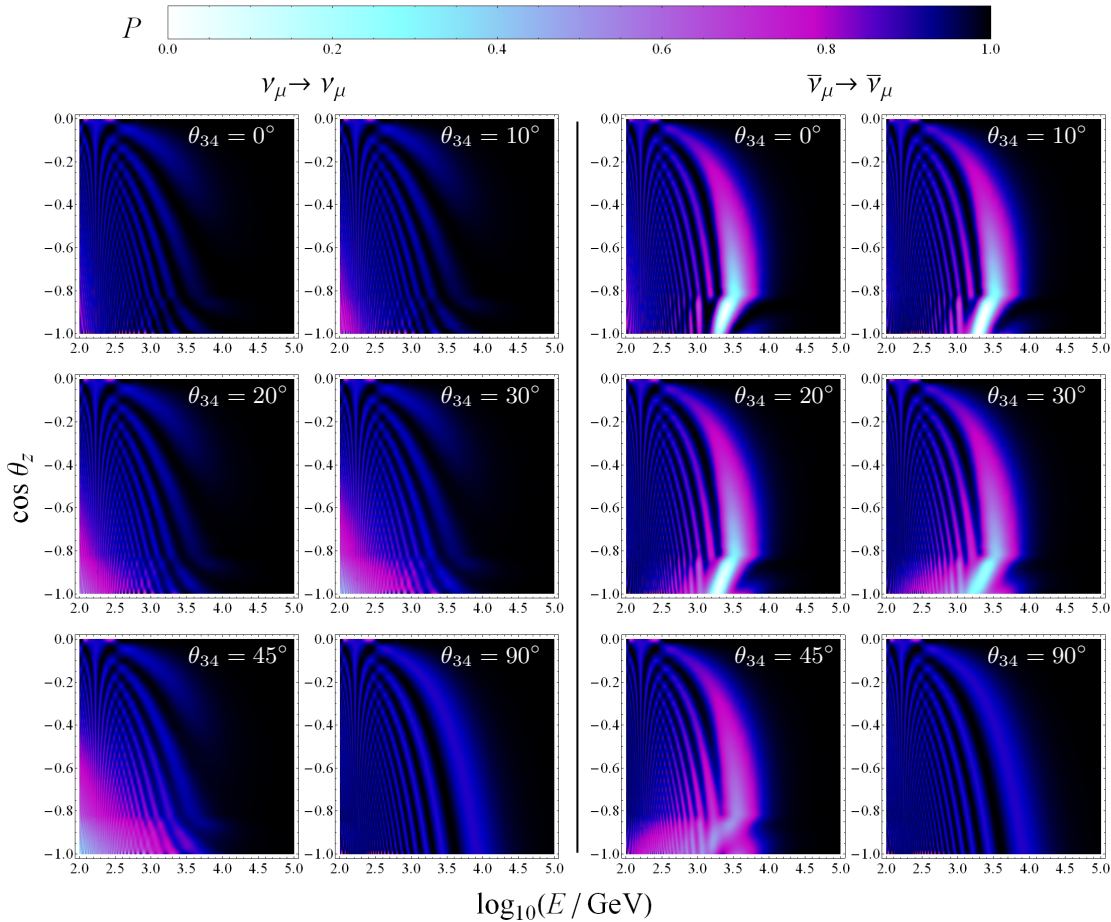

To further verify the importance of $\theta_{34}$, we generate numerical plots without any approximation.
We set $\Delta m_{21}^{2}=7.5\times10^{-5}\,{\rm eV}^{2}$, $\Delta m_{31}^{2}=2.4\times10^{-3}\,{\rm eV}^{2}$,
$\theta_{23}=45{}^{\circ},\thinspace\theta_{13}=9{}^{\circ},\thinspace\theta_{12}=34^{\circ}$
and all CP-violating phases are set to zero. The survival probability as
a function of $E$ and the zenith angle $\cos\theta_{z}$ is plotted in 
Fig.\ \ref{fig:The-survival-probablity}\footnote{Oscillagrams for muon-neutrino oscillations 
into sterile neutrinos (though of much lower scale than discussed here) have been 
first given in Ref.\ \cite{Chizhov:1998ug}.}.  The plot with $\theta_{34}=10^{\circ}$ on the right panel (i.e. antineutrino survival probability) shows
that the probability reaches 0 at $\lg(E/{\rm GeV})\simeq3.4$ and $\cos\theta_{z}\simeq-0.9$
even though all active-sterile mixing angles are small. 
This is due
to the matter effect resonance in the $\overline{\nu}_{\mu}$-channel.
When $U_{s4}$ is reduced, we can see that the resonance becomes weaker and for
$U_{s4}=0$ (i.e.\ $\theta_{34}=90^{\circ}$), the resonance completely 
disappears, and the  result is indistinguishable from the vacuum case.
Note that as $\theta_{34}$ increases the effect at $E_{\nu}=10^2$ GeV and $\cos \theta_z=-1$ becomes significant, which was previously pointed out in \cite{Esmaili:2013vza}. \\

Let us recall here that  short baseline disappearance results are essentially 
electron neutrino appearance results, with an oscillation amplitude of 
\begin{equation}\label{eq:lsnd}
\sin^2 2\theta_{\mu e} = \sin^2 2 \theta_{14} \, \sin^2 \theta_{24}\,.
\end{equation}
We will use later the global fit results from Ref.\ \cite{Kopp:2013vaa} 
for $\sin^2 2\theta_{\mu e}$ 
(see Fig.\ 8 therein), which have been obtained by a fit of available 
 appearance and disappearance results. Also used for comparison 
will be fit results to $\sin^2 2\theta_{\mu e}$ from Ref.\ \cite{Gariazzo:2015rra} 
(see Fig.\ 4 therein) that includes $\nu_{\mu}\to\nu_{e}$ ($\overline{\nu}_{\mu}\to\overline{\nu}_{e}$) appearance 
results in combination with various $\nu_{e}$ ($\overline{\nu}_{e}$) and $\nu_{\mu}$ ($\overline{\nu}_{\mu}$) disappearance 
results (excluding the MiniBooNE low energy excess). 
Another fit result is from Ref.\ \cite{Conrad:2012qt} (see Fig.\ 4 therein). 
The data used in those fits is not always the same, as is the treatment 
of the data, so differences arise.
However, the analyses of \cite{Kopp:2013vaa} and \cite{Conrad:2012qt}, using very similar data sets, are giving results in 
approximate agreement with each other.
 Hence statements regarding ruling out sterile neutrino hints will depend on the fit result one compares to. Less differences arise for fit results of only appearance data, and we will compare to the results from Ref.\ \cite{Kopp:2013vaa} on 
$\nu_{\mu}\to\nu_{e}$ ($\overline{\nu}_{\mu}\to\overline{\nu}_{e}$) appearance data. 
Let us note that the LSND results are crucial 
for the hints for sterile neutrinos, 
excluding them from global fits reduces the significance dramatically \cite{Gariazzo:2015rra}. 
Another bound of interest is from Super-Kamiokande \cite{Abe:2014gda}, which found 
$|U_{\mu4}|^2 <0.054$ at 99\% C.L., though with assuming $U_{e4}=0$ 
and $\Delta m^2_{41} > 0.1$ eV$^2$, and a limit on $|U_{e4}|^2$ of about 0.09. 
Finally, we should mention the 90\% C.L.\ constraint $\theta_{34} \le 25^\circ$, obtained from an analysis 
of muon neutrino disappearance in the MINOS experiment \cite{Adamson:2011ku}. 
Regarding electron (anti)neutrino disappearance results, severe tension with various 
appearance results exists \cite{Kopp:2013vaa,Conrad:2012qt,Gariazzo:2015rra}, resembling situations 
in which inconsistent data sets are combined. Anyway, later we will 
often take the example values $\theta_{14} = 4^\circ$ and $\theta_{14} = 10^\circ$, which are 
compatible with the $3\sigma$ ranges of a most recent global appearance and disappearance fit 
from Ref.\ \cite{Gariazzo:2015rra}. \\

Note that from Eq.\ (\ref{eq:0908-3}) one can show that the 
presence of matter effects will in general make muon survival probabilities depend on 
$U_{\mu 4}$ and $U_{\tau 4}$, hence on 
$\theta_{14}, \theta_{24}$ and $\theta_{34}$ \cite{Esmaili:2013vza}. 
Indeed, assuming $U_{e4}=0$ 
(this matrix element has little influence on the final result, as we have essentially 
a two flavor oscillation case) and following the same calculation 
as the one leading to Eq.\ (\ref{eq:0908-8}), leads to 
\begin{equation}
S_{\mu\mu}=\frac{\mu_{4}^{2}e^{-\frac{1}{2}it(A\kappa+C_{2}+1)}\left[e^{C_{2}it}\left(A\kappa+C_{2}-C_{3}\right)-A\kappa+C_{2}+C_{3}\right]+2C_{2}\tau_{4}^{2}}{2C_{2}\left(\mu_{4}^{2}+\tau_{4}^{2}\right)},
\label{eq:0916}
\end{equation}
where 
$
C_{2}=\sqrt{1+A\kappa\left(2-4(\mu_{4}^{2}+\tau_{4}^{2})+A\kappa\right)}
$ and 
$C_{3}=2\mu_{4}^{2}+2\tau_{4}^{2}-1.$
The muon-neutrino survival probability is therefore to good precision a function of 
$|U_{\mu 4}|^2/|U_{\tau 4}|^2 = \tan^2 \theta_{24}/\sin^2 \theta_{34}$. Again, we see that the comparison 
of IceCube atmospheric neutrino results with 
short baseline disappearance experiments depends on $\theta_{34}$. 

\section{\label{sec:atm} Neutrino event numbers in IceCube}

The neutrino event numbers depend on the neutrino energy $E$ and
the zenith angle $c_{z}\equiv\cos\theta_{z}$. It can be computed
via 
\begin{equation}
\frac{\partial^{2}N}{\partial E\partial c_{z}}(E,c_{z})=2\pi T\left[A_{eff}\Phi P+\bar{A}_{eff}\bar{\Phi}\bar{P}\right],\label{eq:0819}
\end{equation}
where $A_{eff}$  ($\bar{A}_{eff}$) is the effective area of IceCube
for $\nu_{\mu}$ ($\overline{\nu}_{\mu}$), $\Phi$ ($\bar{\Phi}$)
is the flux of $\nu_{\mu}$ ($\overline{\nu}_{\mu}$) and $P$ ($\bar{P}$)
is the survival probability of $\nu_{\mu}$ ($\overline{\nu}_{\mu}$).
All three quantities ($A_{eff},\thinspace\Phi,\thinspace P$)
are functions of the neutrino energy $E$ and the zenith angle $c_{z}$.
For the IceCube-59 data, $T=348.1$ days \cite{Aartsen:2013eka}. The
factor 2$\pi$ is due to integration over the azimuthal angle. 

For $\Phi$ and $\bar{\Phi}$, we use the data from Ref.\ \cite{Honda:2006qj}.
Since the flux has been computed only up to $10^{4}$ GeV while a small
part of events in the IceCube-59 data have energies above $10^{4}$ GeV
(most events are in the energy range from $10^{2}$ GeV to $10^{4}$ GeV),
we need to extrapolate the data to $10^{6}$ GeV to cover the full
data. The final result should be insensitive to the extrapolation
because the high energy part has little contribution to the total event
number. 

The effective area can be extracted from \cite{Aartsen:2013eka}, where in 
Fig.\ 1 the simulation results for the event numbers as functions of energy 
and of zenith angle (without neutrino oscillation) are shown for the conventional 
atmospheric neutrino flux. So $A_{eff}\Phi$ 
can be obtained from that figure and $A_{eff}$ can be extracted, provided 
that $\Phi$ is known. In practice, a more detailed procedure is adopted by us 
to take into account  the difference between $\nu$ and $\overline{\nu}$
and the zenith angle dependence of $A_{eff}$: we 
assume an energy-dependent ratio of $A_{eff}$ to $\bar{A}_{eff}$,
\begin{equation}
\bar{A}_{eff}(E,c_{z})=\lambda(E)A_{eff}(E,c_{z})\,,\label{eq:0909-4}
\end{equation}
where the ratio $\lambda(E)$ can be taken from Fig.\ 2 of Ref.\ \cite{Abbasi:2009nfa}. 
We also assume that the dependence of $A_{eff}$ on the zenith angle
is mainly due to the detection efficiency of photons generated by the muon
tracks, since the cross section $\sigma(E)$ of neutrinos with nuclei
of water molecules should only depend on $E$. Under this assumption,
we have 
\begin{equation}
A_{eff}=\sigma(E)Nf(c_{z})\,,\label{eq:0909-3}
\end{equation}
where $N$ is the number of water molecules and $f(c_{z})$ is the
detection efficiency. So without neutrino oscillation, Eq.\ (\ref{eq:0819})
becomes 
\begin{equation}
\frac{\partial^{2}N}{\partial E\partial c_{z}}(E,c_{z})=f(c_{z})g(E)[\Phi(E,C_{z})+\lambda(E)\bar{\Phi}(E,C_{z})]\,,\label{eq:0819-2}
\end{equation}
where $g(E)\equiv2\pi T\sigma(E)N$. In Eq.\ (\ref{eq:0819-2}) only
$f(c_{z})$ and $g(E)$ are unknown functions to be determined and there are
two curves in Fig.\ 1 of Ref.\ \cite{Aartsen:2013eka} for 
$\int\frac{\partial^{2}N}{\partial E\partial c_{z}}dE$ 
and $\int\frac{\partial^{2}N}{\partial E\partial c_{z}}dc_{z}$, correspondingly.
So we can solve for $f$ and $g$ from the two curves to obtain $A_{eff}$
and $\bar{A}_{eff}$. 
Note that the detection efficiency function of Cherenkov
photons $f$ in Eq.\ (\ref{eq:0909-3}) (which should mainly depend on $c_{z}$ because
it is essentially a geometrical effect) may also have weak dependence
on energy. However, events in IceCube are not uniformly distributed
from $10^{2}$ GeV to $10^{6}$ GeV, but are rather concentrated around $10^{3}$ GeV.
Therefore, the effective integration region of $\int\frac{\partial^{2}N}{\partial E\partial c_{z}}dE$
 is very narrow and even if $f$ depends 
weakly on $E$, only those $f$-values around $10^{3}$ GeV are important. 

In the actual measurement $E$ can only be partially reconstructed
from the muon track; only a lower bound on $E$ can be obtained 
from the truncated energy loss of the muon (see  Fig.\ 4 in Ref.\ \cite{Aartsen:2013eka}). 
Besides, the correlation of the true energy $E$ of the neutrino and the truncated
energy loss of the muon is very difficult for us to handle. 
So in this paper we will not use the energy spectrum information of the data and simply
integrate over $E$ in Eq.\ (\ref{eq:0819}), though this will somewhat reduce
the sensitivity on sterile neutrinos. A full analysis involving
the energy spectrum information should be implemented by the IceCube
collaboration. 

\begin{figure}[t]
\centering

\includegraphics[width=8cm]{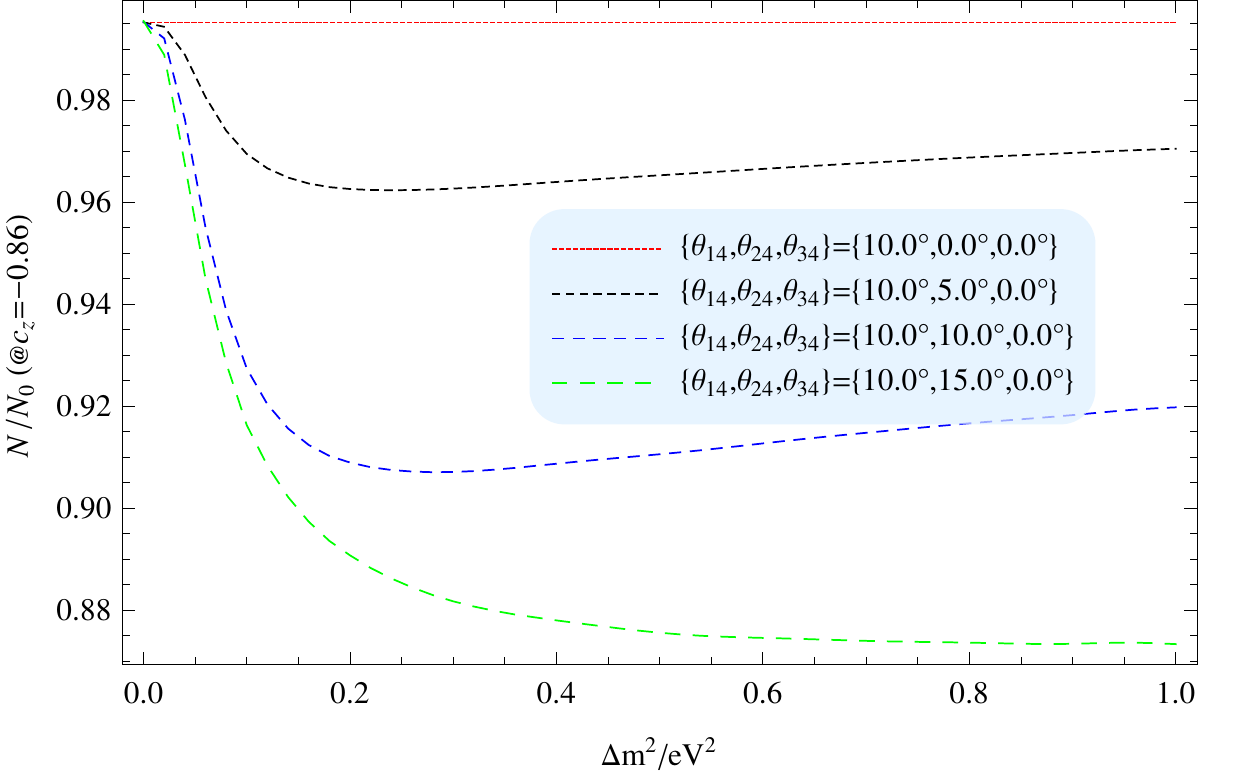}\includegraphics[width=8cm]{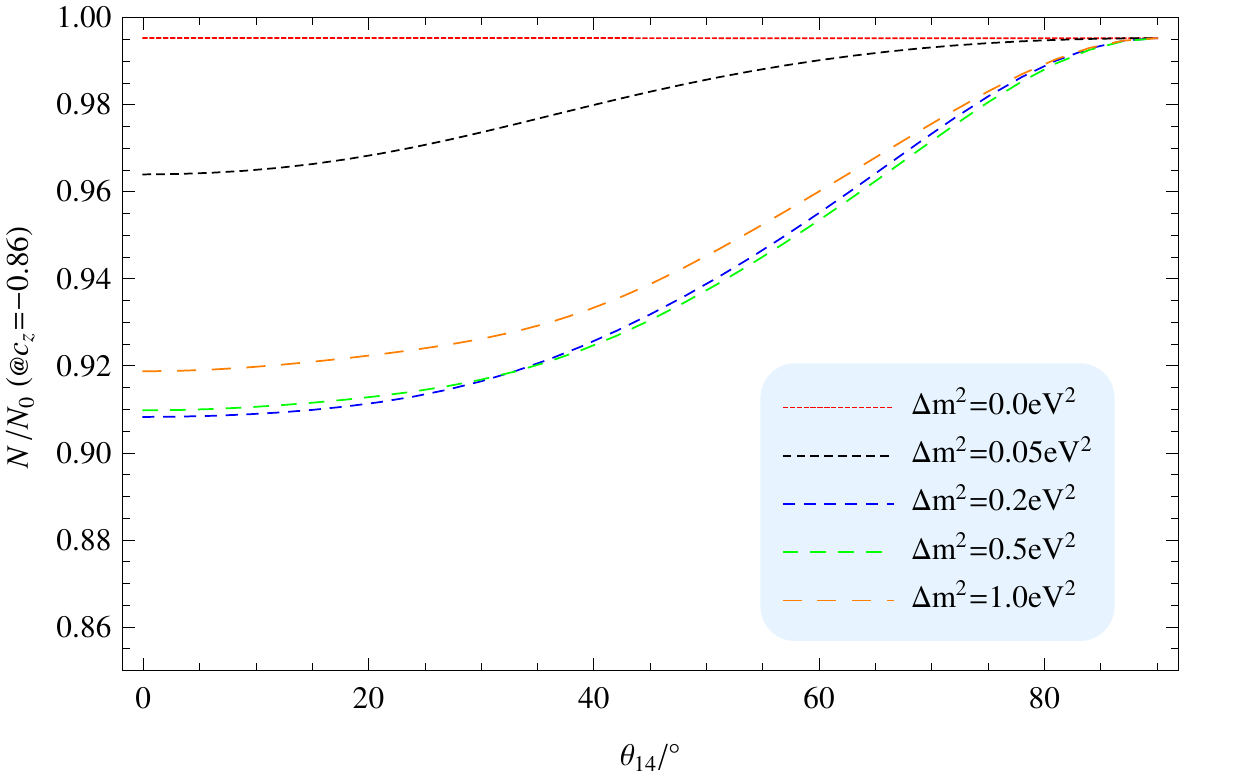}

\includegraphics[width=8cm]{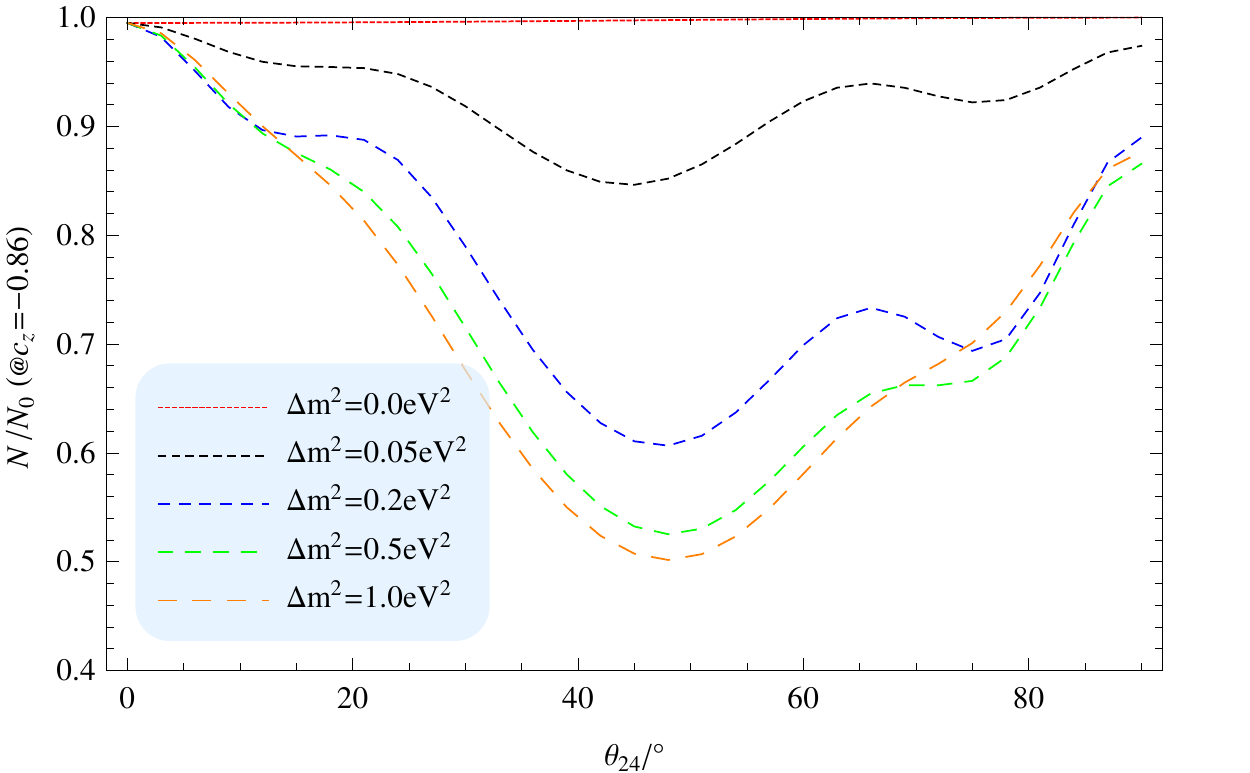}\includegraphics[width=8cm]{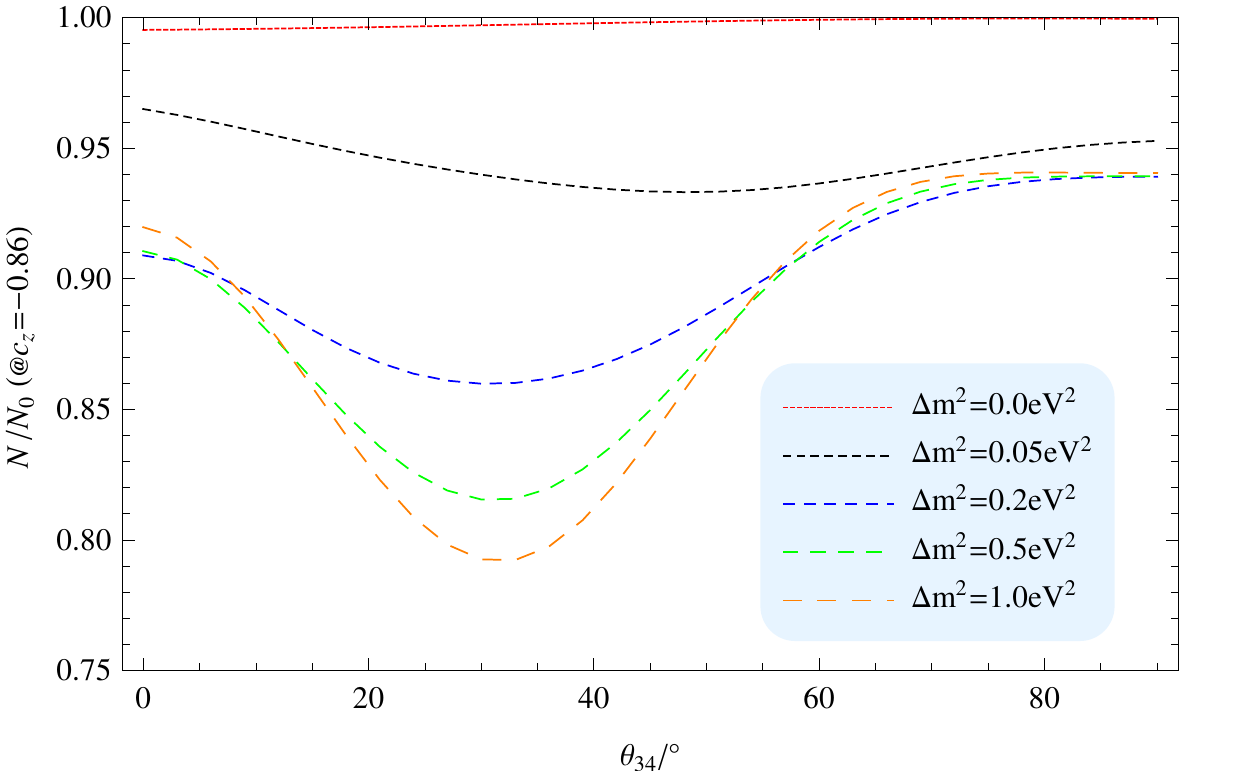}\caption{\label{fig:curves}
The ratio of event number $N$ and the unoscillated event number $N_0$ as function of the relevant sterile neutrino parameters. 
We have chosen the illustrative value $\cos \theta_z = -0.86$; 
the parameters $\theta_{14}$ and $\theta_{24}$, if not specified in the plots, are set at $10^\circ$. But for $\theta_{34}$ it is $0^\circ$.
}
\end{figure}

\begin{figure}[h]
\centering

\includegraphics[height=9cm]{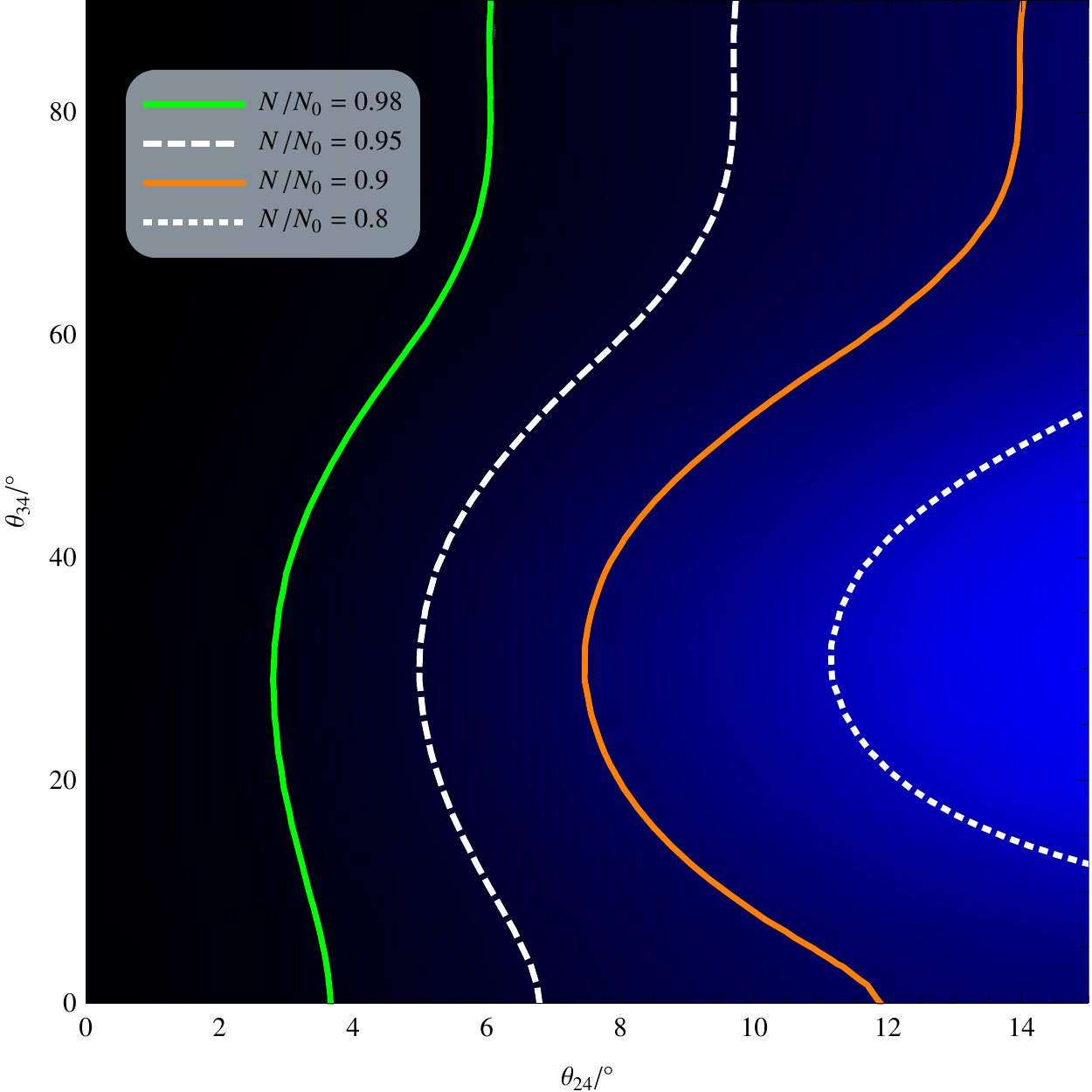}\caption{\label{fig:2434}
The ratio of event number $N$ and the unoscillated event number $N_0$ at $\cos\theta_{z}=-0.86$ in the
$\theta_{24}-\theta_{34}$ plane for 
$\Delta m^{2}=0.5\,{\rm eV}^{2}$ and $\theta_{14}=0^\circ$. 
}
\end{figure}

Hence, we integrate over $E$ and compute the event number in each small
$c_{z}$-bin,
\begin{equation}
N(c_{z})=2\pi T\Delta_{c_{z}}\intop E\ln(10)\left[A_{eff}(\lg E,c_{z})\Phi(\lg E,c_{z})P(\lg E,c_{z})+(\nu\rightarrow\overline{\nu})\right]d\lg E\,,\label{eq:0819-1}
\end{equation}
where $\Delta_{c_{z}}$ is the width of $\cos\theta_{z}$-bins, equal
to $1/25$ for 25 bins and 
\begin{equation}
\lg E\equiv\log_{10}(E/{\rm GeV})\,.\label{eq:0821}
\end{equation}
Note that though reconstructed zenith angles are not true zenith angles of neutrinos, for energies relevant to our analysis and track events, the difference is much smaller than the bin width \cite{Aartsen:2013eka}, and thus 
we will not consider the difference between reconstructed and true zenith angles in our analysis.
For later use, we also define the no-oscillation event number $N_{0}$,
\begin{equation}
N_{0}(c_{z})=2\pi T\Delta_{c_{z}}\intop E\ln(10)\left[A_{eff}(\lg E,c_{z})\Phi(\lg E,c_{z})+(\nu\rightarrow\overline{\nu})\right]d\lg E\,.\label{eq:0909-5}
\end{equation}
Although the total event number is overall reduced due to neutrino disappearance, 
in practice this effect is not useful to constrain sterile neutrino parameters. 
The reason is that the fluxes $\Phi$ and $\bar \Phi$ have a large uncertainty, e.g.\ 
about $25\%$ at $10^{3}$ GeV, indicated in Fig.\ 
11 in \cite{Honda:2006qj}. Compared to 
the statistical uncertainty (about $3\%$) and the systematic uncertainty
(about $4\%$ \cite{ICRC}) in each $c_{z}$-bin, the large uncertainty in the
normalization factor implies the flux can be almost freely renormalized. 

The main observable effect caused by a sterile neutrino in IceCube
is tilting the zenith angle ($\cos\theta_{z}$) distribution of events, 
i.e.\ the smaller $c_{z}$ the more the event numbers are suppressed \cite{Esmaili:2013vza}. 
The existence of sterile 
neutrinos causes disappearance for atmospheric neutrinos going through
Earth. For a very small $|c_{z}|$ the corresponding oscillation 
baseline is very short hence neutrinos do not have enough time to oscillate
before they arrive at the detector. So the survival probability is 
always very close to 1 if $c_{z}$ is small enough, no matter how large
the mixing angles are\footnote{Note that for large $\Delta m^{2}$,
 neutrino oscillation may still happen in the horizontal direction
(i.e.\ $c_{z}=0$); for this qualitative discussion we ignore such aspects, the 
numerical analysis takes those effects into account.}. 
For a large $|c_{z}|$, neutrinos may have propagated over 
enough baseline to oscillate and thus the survival probability could be
low. They may also experience several oscillatory periods before they
arrive and the survival probability would be a value between the minimum and $1$, 
depending on energy and zenith angle. This qualitative 
analysis can be verified from Fig.\ \ref{fig:The-survival-probablity}, 
where $P$ is always close to 1 at $c_{z}\simeq0$. For a large $|c_{z}|$,
$P$ can be relatively small or still large (close to $1$), depending
on $E$. Note that Eq.\ (\ref{eq:0819-1}) is an integral over $E$ 
so only the average value (roughly) of $P$ is of importance.
In this sense, we can say that the disappearance signal is stronger
at larger $|c_{z}|$ and weaker at small $|c_{z}|$. Therefore, the
sterile neutrino signal in the zenith angle distribution is mainly
a tilting effect.
To obtain a qualitative understanding of the sensitivity on sterile neutrino
parameters, we plot in Fig.\ \ref{fig:curves} for illustration 
$N/N_{0}$ at $c_{z}=-0.86$ for the various sterile neutrino parameters 
($\Delta m^{2},\theta_{14},\theta_{24},\theta_{34}$),  
i.e.\ the ratio of events for 
the oscillated and unoscillated case at a large zenith angle. 
The more this quantity deviates from $1$, the more the zenith angle distribution
tilts. The upper left $\Delta m^{2}$-plot shows that $N/N_{0}$ drops
down quickly from $1$ when $\Delta m^{2}$ goes from $0$ to $0.1\,{\rm eV}^{2}$.
This implies that the sensitivity of IceCube on sterile neutrinos
depends on $\Delta m^{2}$ significantly at this range. The upper right $\theta_{14}$-plot
shows that $N/N_{0}$ changes very little for $\theta_{14}\in[0^{\circ},20^{\circ}]$
which means IceCube is insensitive to small $\theta_{14}$ (note that
large $\theta_{14}$ has been excluded by reactor neutrino experiments).
There is almost no difference between $\theta_{14}=0^{\circ}$ and
$10^{\circ}$ for IceCube. The angle $\theta_{24}$ is the most sensitive parameter as shown in 
the lower left $\theta_{24}$-plot. For (too) large $\theta_{24}$ such as 
$40^{\circ}$, $N/N_{0}$ could drop to $0.5$. 
For small $\theta_{24}$, we should include $\theta_{34}$ 
in the sensitivity analysis because the matter effect resonance is sensitive
to the ratio $U_{\mu4}/U_{\tau4} = \tan \theta_{24}/\sin \theta_{34}$, see Eq.\ (\ref{eq:0916}).  
Moreover, the important matter effect could decouple for large $\theta_{34}$, as
we have discussed in Section \ref{sec:atm}. 
We therefore plot $N/N_{0}$ in the $\theta_{24}-\theta_{34}$ plane in Fig.\ \ref{fig:2434}, 
where we can see that for $\theta_{34}=20^{\circ}$ 
even a small $\theta_{24}$ value ($8^{\circ}$) can make $N/N_{0}$ drop to $0.9$. 
Note that when $\theta_{34}$ is close to $90^{\circ}$
then as we have mentioned in the analytic discussion, $\nu_{\mu}$ 
($\overline{\nu}_{\mu}$) neutrinos would oscillate as if they are
in vacuum. So generally speaking, the signal of sterile neutrinos
for $\theta_{34}=90^{\circ}$ is weaker than for other values.

\begin{figure}[t]
\centering
\includegraphics[width=12cm]{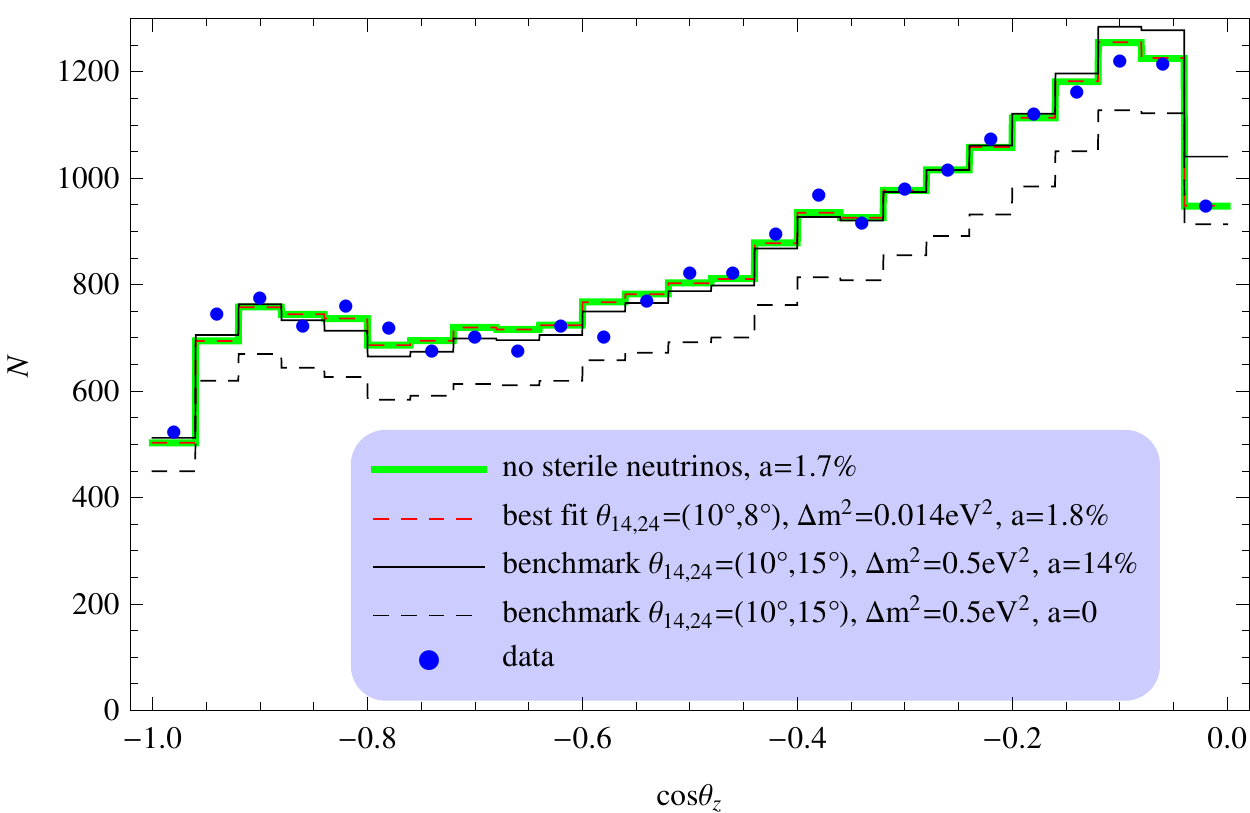}
\caption{\label{fig:distributions}
Event numbers versus zenith angle $\cos \theta_{z}$  for several cases. The atmospheric neutrino flux is normalized by a factor of $1+a$, see Eq.\ (\ref{eq:0820}). 
The data can be fitted very well without any sterile neutrino contribution. 
}
\end{figure}


\begin{figure}[hp]
\centering
\begin{minipage}{16cm}
	\begin{minipage}{0.5cm}
	\centering
	\begin{turn}{90}$\log_{10}(\Delta m^{2}/{\rm eV}^{2})$\end{turn}
	\end{minipage}
	\begin{minipage}{12.5cm}
	\centering
	\begin{overpic}[width=6cm]{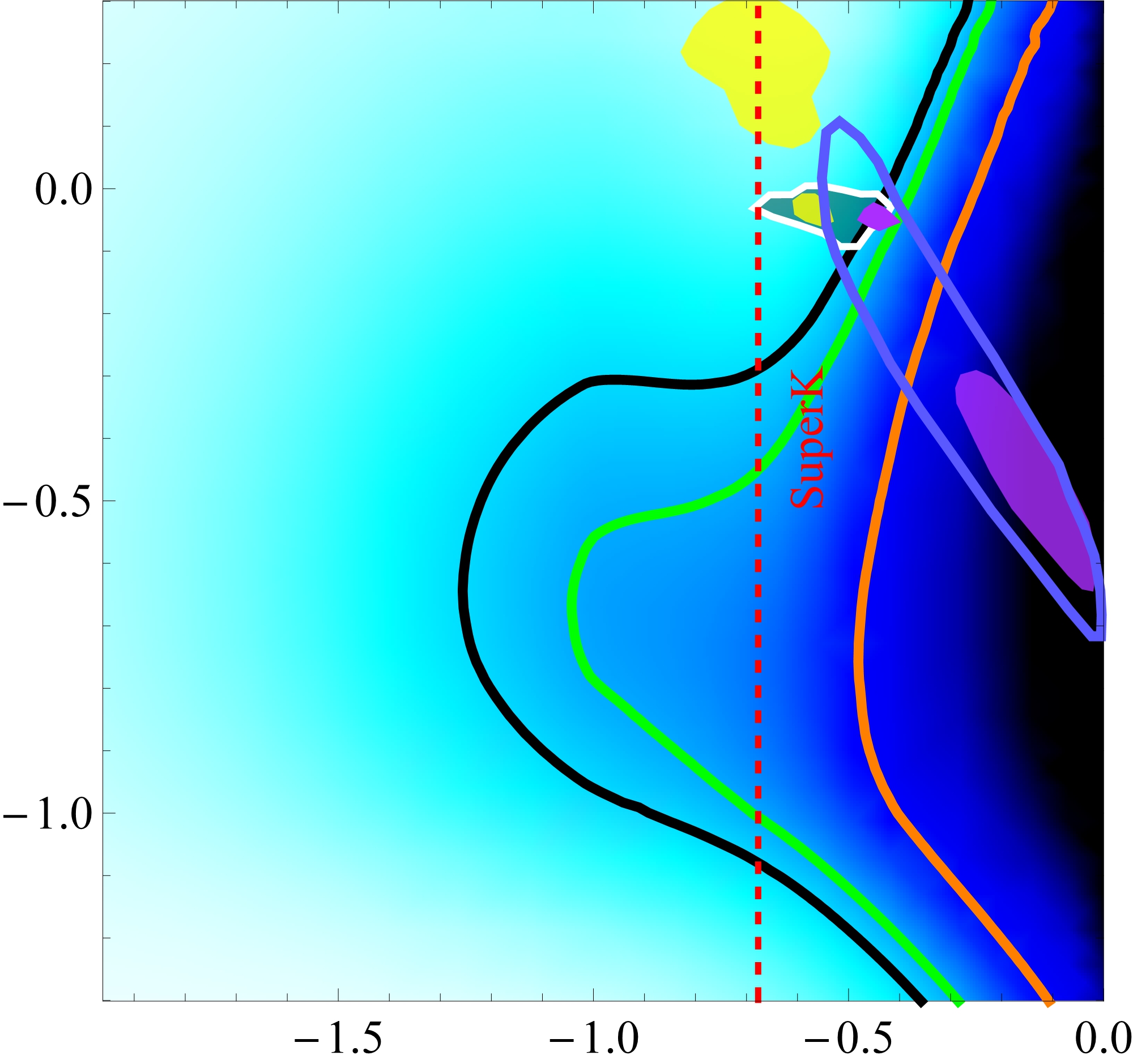}
	\put (10,85) {$\theta_{34}=0^{\circ}$}
	\put (10,80) {$\theta_{14}=10^{\circ}$}
	\end{overpic}\quad
	\begin{overpic}[width=6cm]{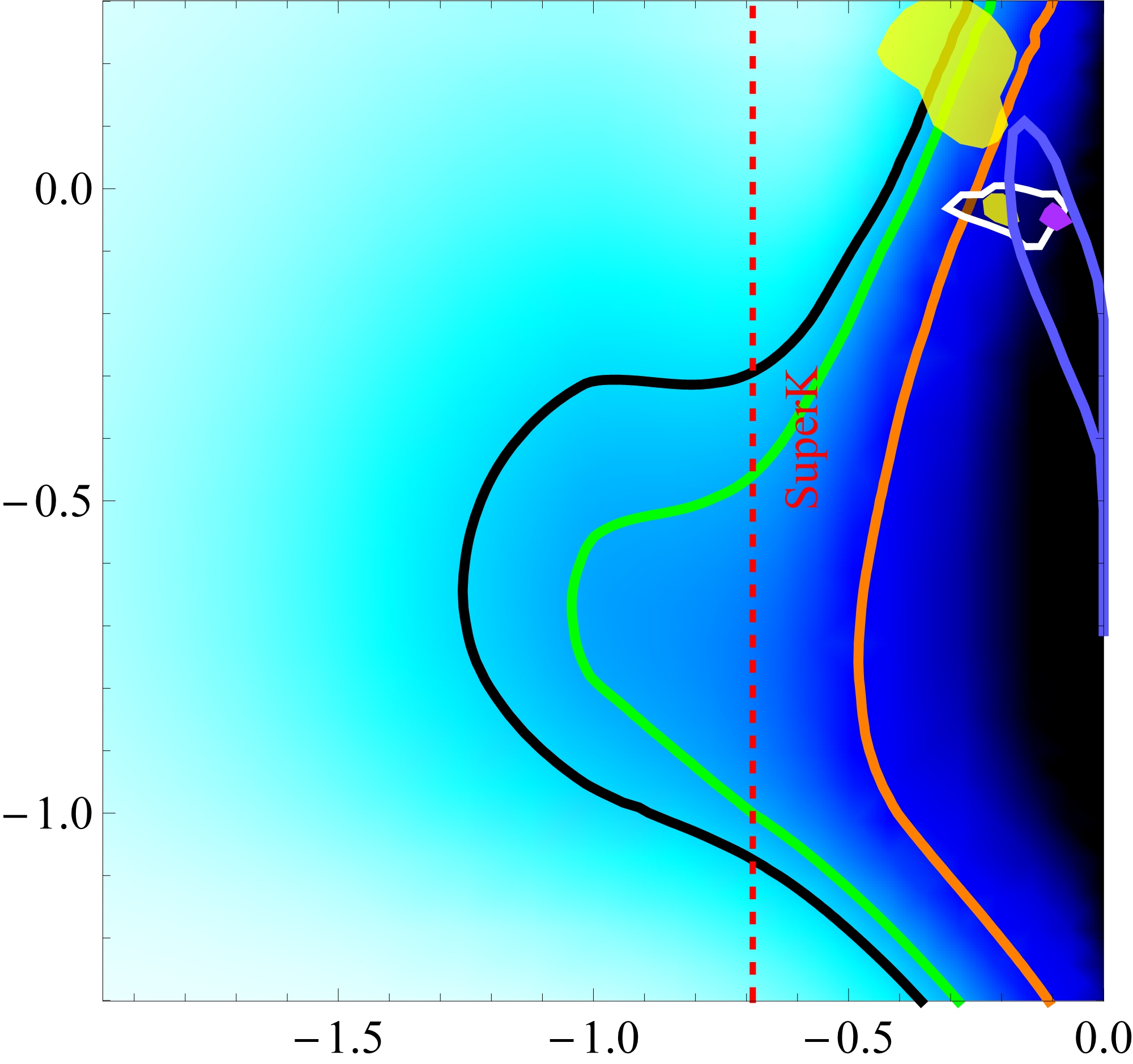}
	\put (10,85) {$\theta_{34}=0^{\circ}$}
		\put (10,80) {$\theta_{14}=4^{\circ}$}
	\end{overpic}\\
	\begin{overpic}[width=6cm]{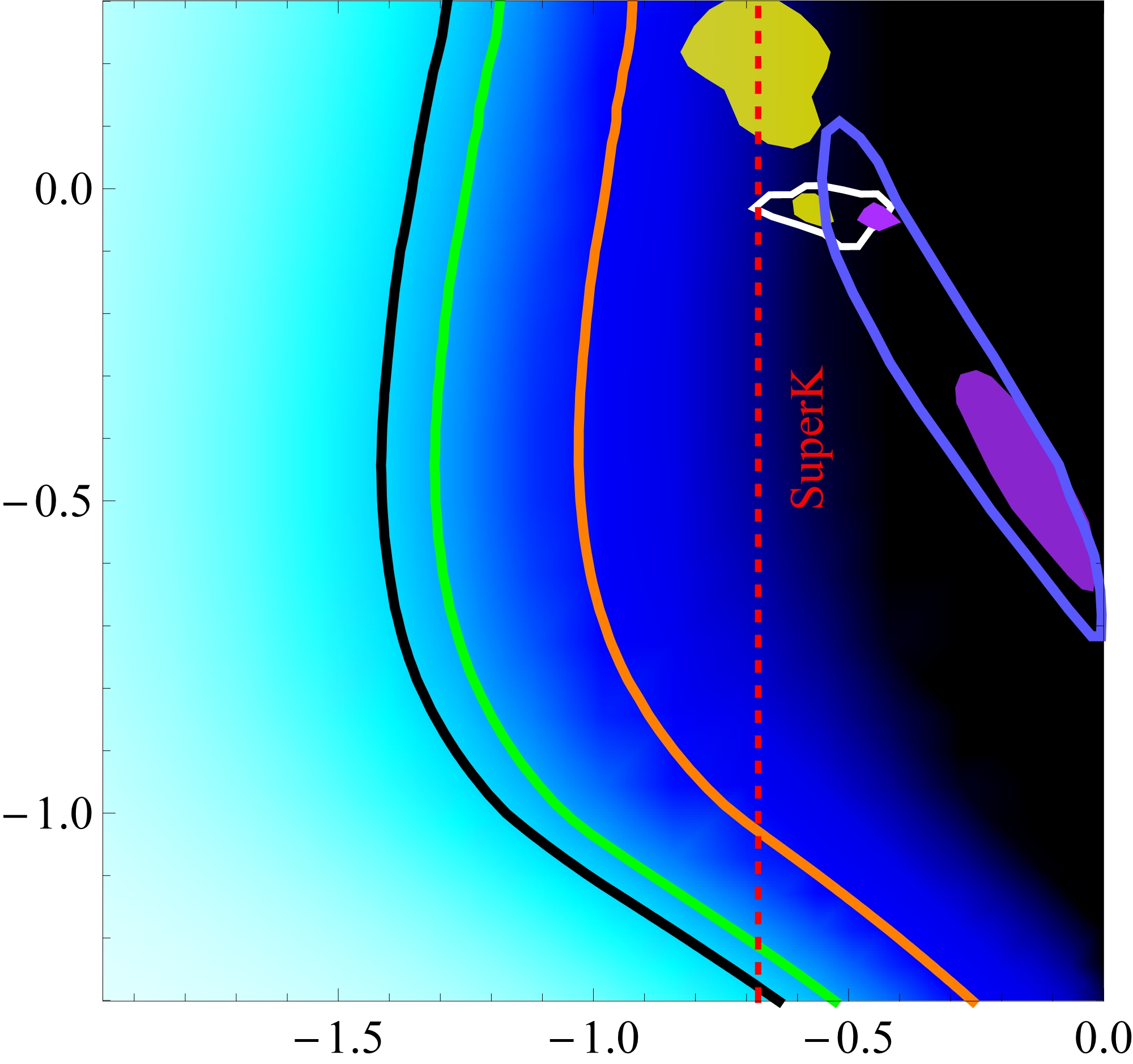}
	\put (10,85) {$\theta_{34}=10^{\circ}$}
		\put (10,80) {$\theta_{14}=10^{\circ}$}
	\end{overpic}\quad
	\begin{overpic}[width=6cm]{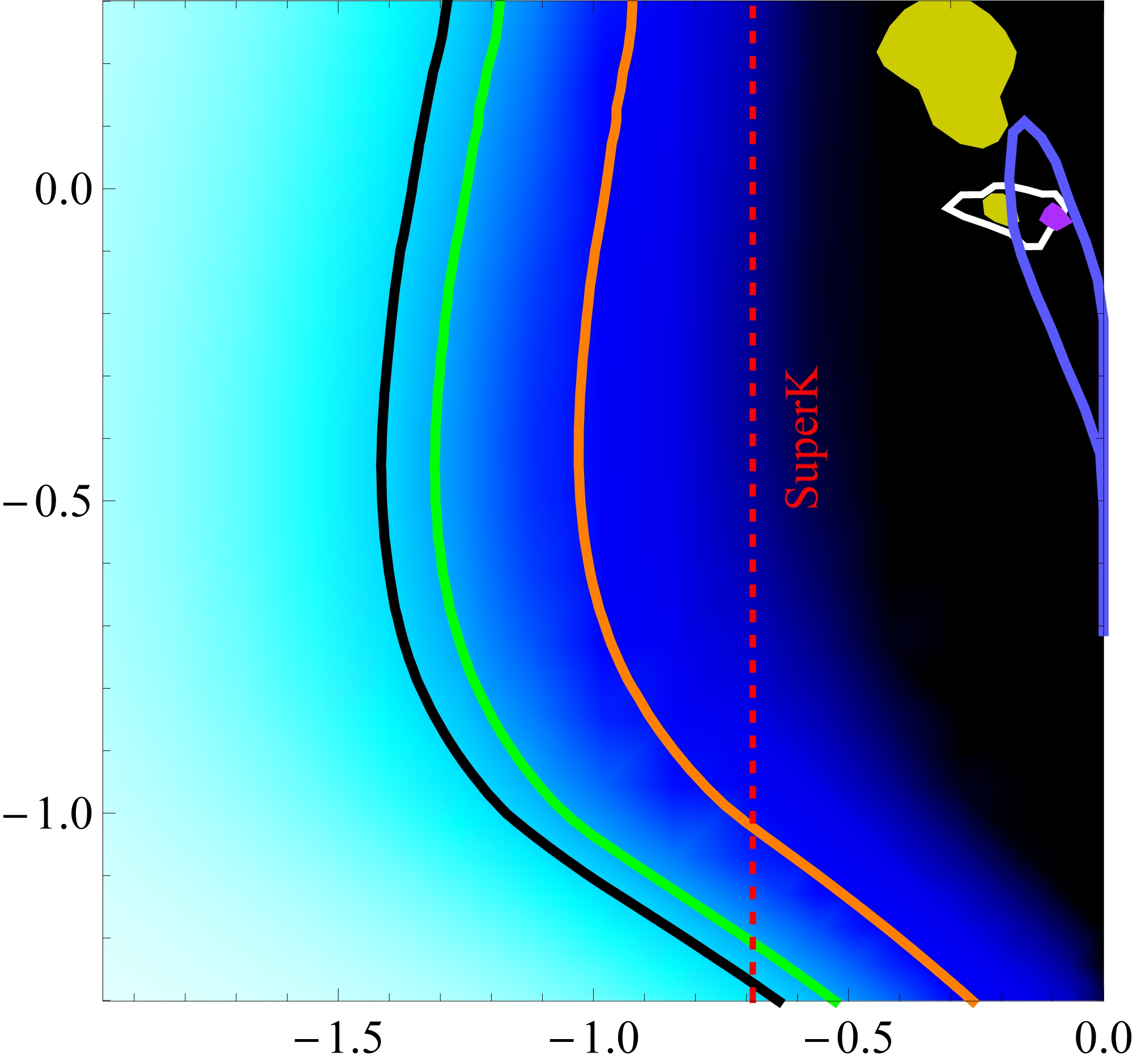}
	\put (10,85) {$\theta_{34}=10^{\circ}$}
		\put (10,80) {$\theta_{14}=4^{\circ}$}
	\end{overpic}\\
	\begin{overpic}[width=6cm]{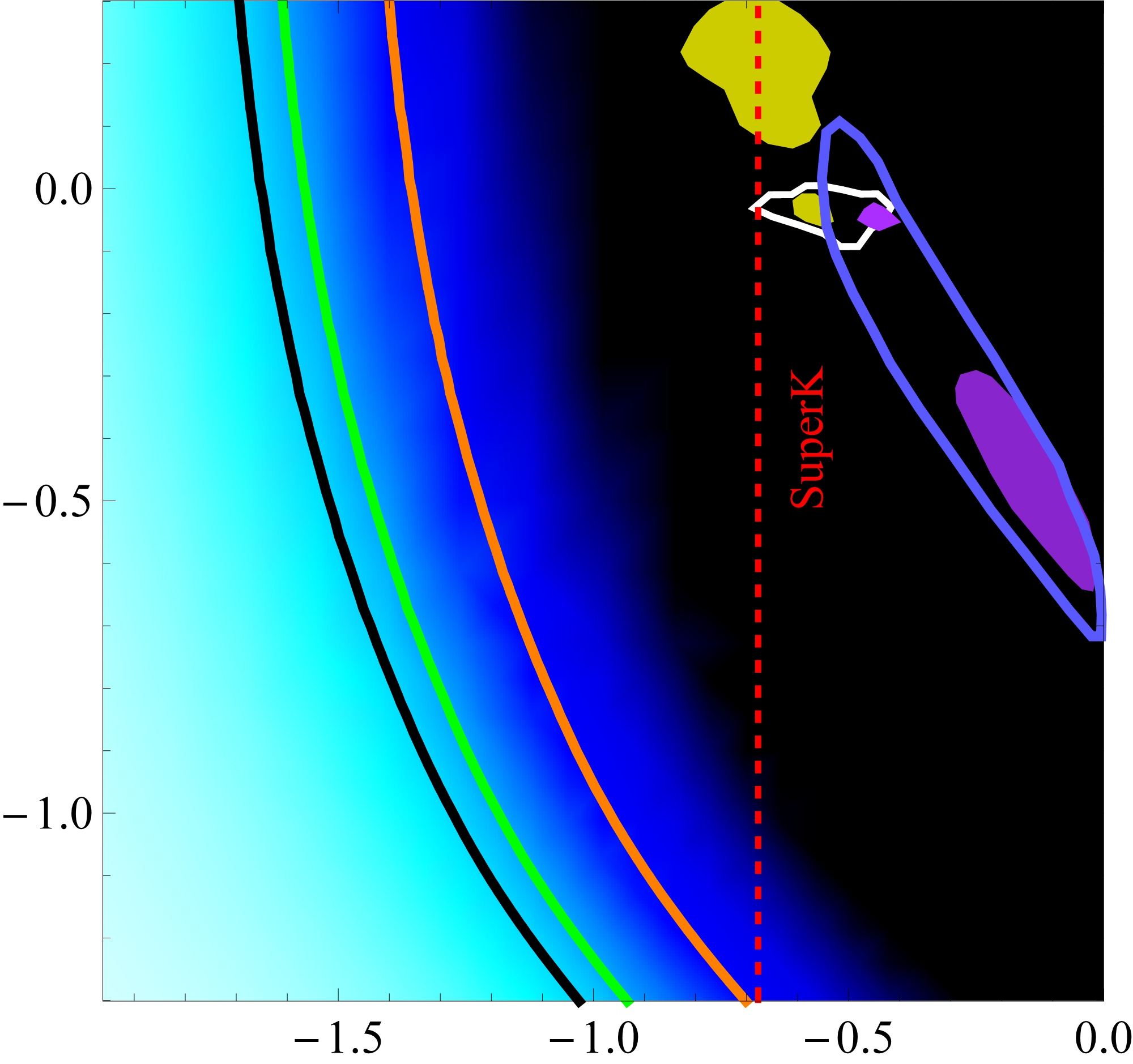}
	\put (10,15) {$\theta_{34}=30^{\circ}$}
		\put (10,10) {$\theta_{14}=10^{\circ}$}
	\end{overpic}\quad
	\begin{overpic}[width=6cm]{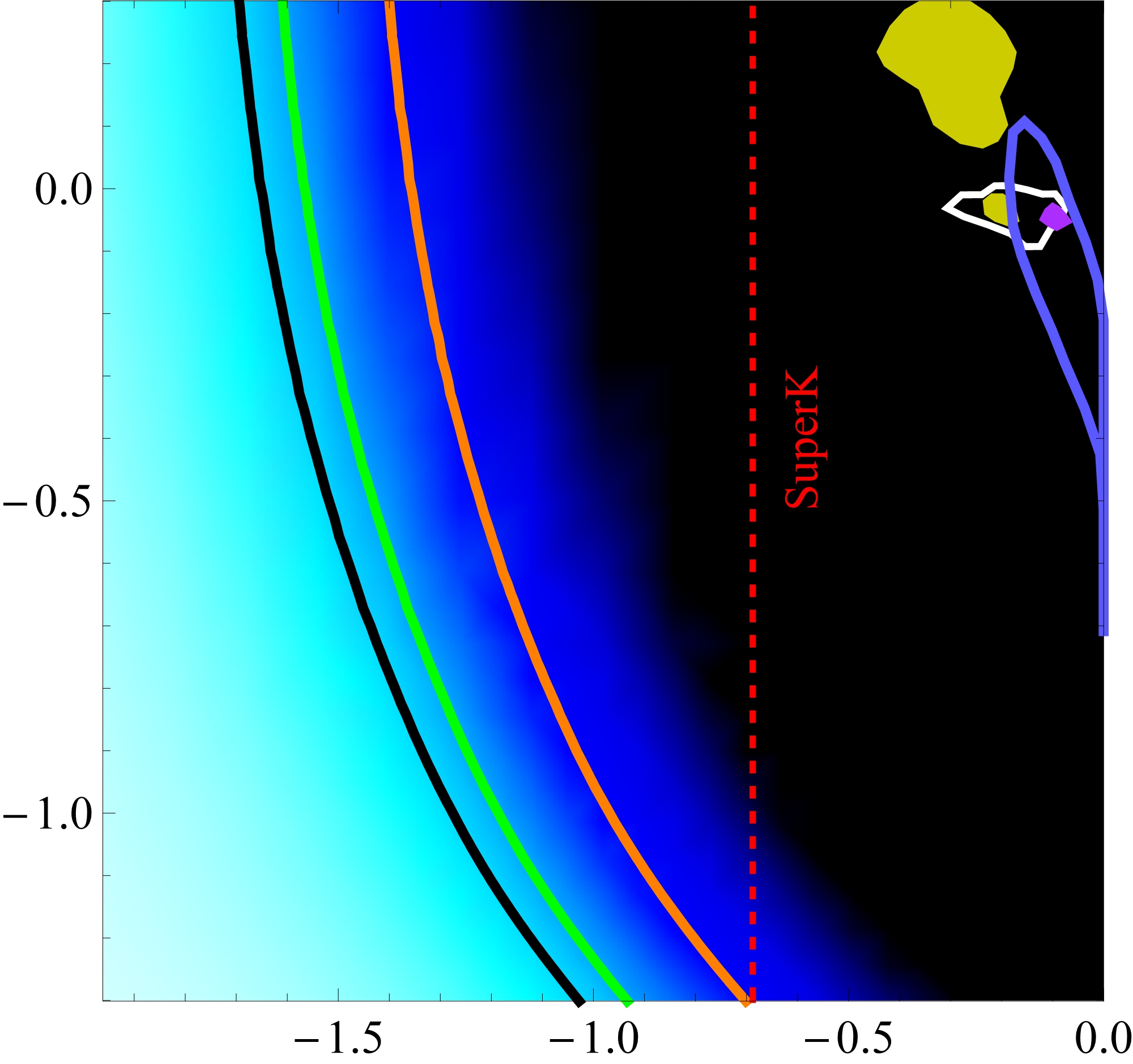}
	\put (10,15) {$\theta_{34}=30^{\circ}$}
		\put (10,10) {$\theta_{14}=4^{\circ}$}
	\end{overpic}\\
	\begin{overpic}[width=6cm]{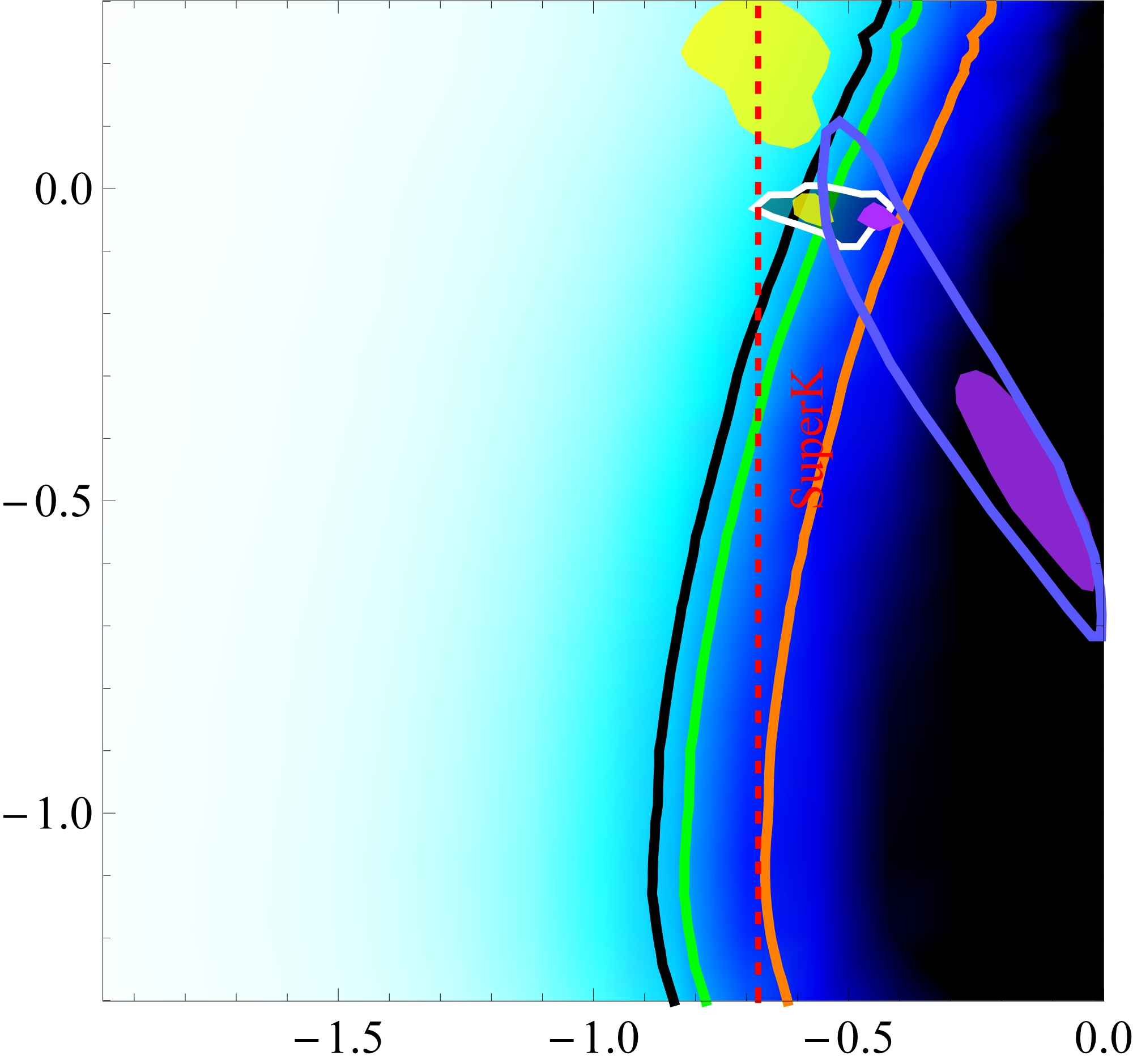}
	\put (10,85) {$\theta_{34}=90^{\circ}$}
		\put (10,80) {$\theta_{14}=10^{\circ}$}
	\end{overpic}\quad
	\begin{overpic}[width=6cm]{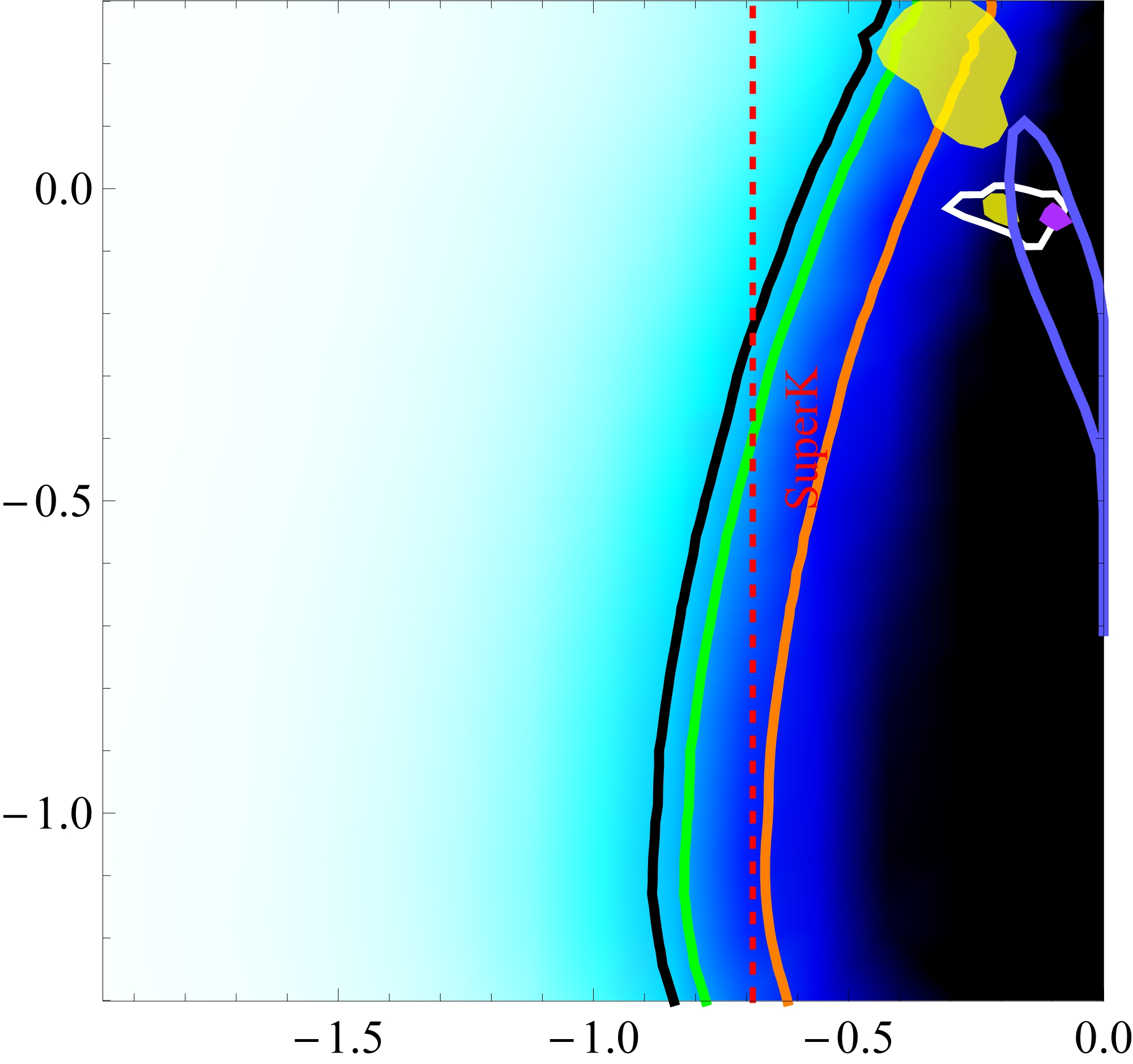}
	\put (10,85) {$\theta_{34}=90^{\circ}$}
		\put (10,80) {$\theta_{14}=4^{\circ}$}
	\end{overpic}\\
	$\log_{10}\sin^{2}2\theta_{24}$    
	\end{minipage}
	\begin{minipage}{2cm}
	\includegraphics[width=1cm]{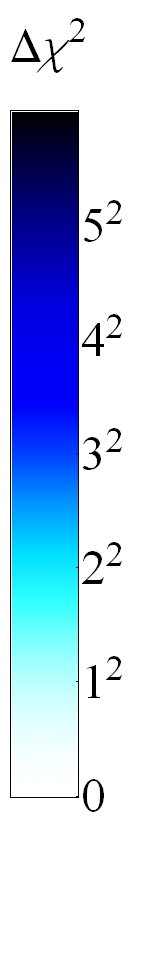}\\
	\includegraphics[width=1.9cm]{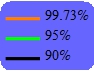}
	\end{minipage}
\end{minipage} 
\caption{\label{fig:Exclusion-bound}Exclusion bounds for $(\Delta m^{2},\sin^{2}2\theta_{24})$.
In these plots, $\theta_{34}$ is fixed at (from top to bottom) 
$0^{\circ}$, $10^{\circ}$, $30^{\circ}$ and $90^{\circ}$; 
$\theta_{14}$ at $10{}^{\circ}$ (left)
or $4^{\circ}$ (right). Also shown are 99\% C.L.\ fit results from the world's appearance and disappearance 
data in purple (from Kopp \emph{et~al.\ }\cite{Kopp:2013vaa}), 
in yellow (from Giunti \emph{et~al.\ }\cite{Gariazzo:2015rra}) and in 
the white contour (from  Conrad \emph{et~al.\ }\cite{Conrad:2012qt}). The blue contour line is the result of an 
appearance data only fit from \cite{Kopp:2013vaa}.}
\end{figure}
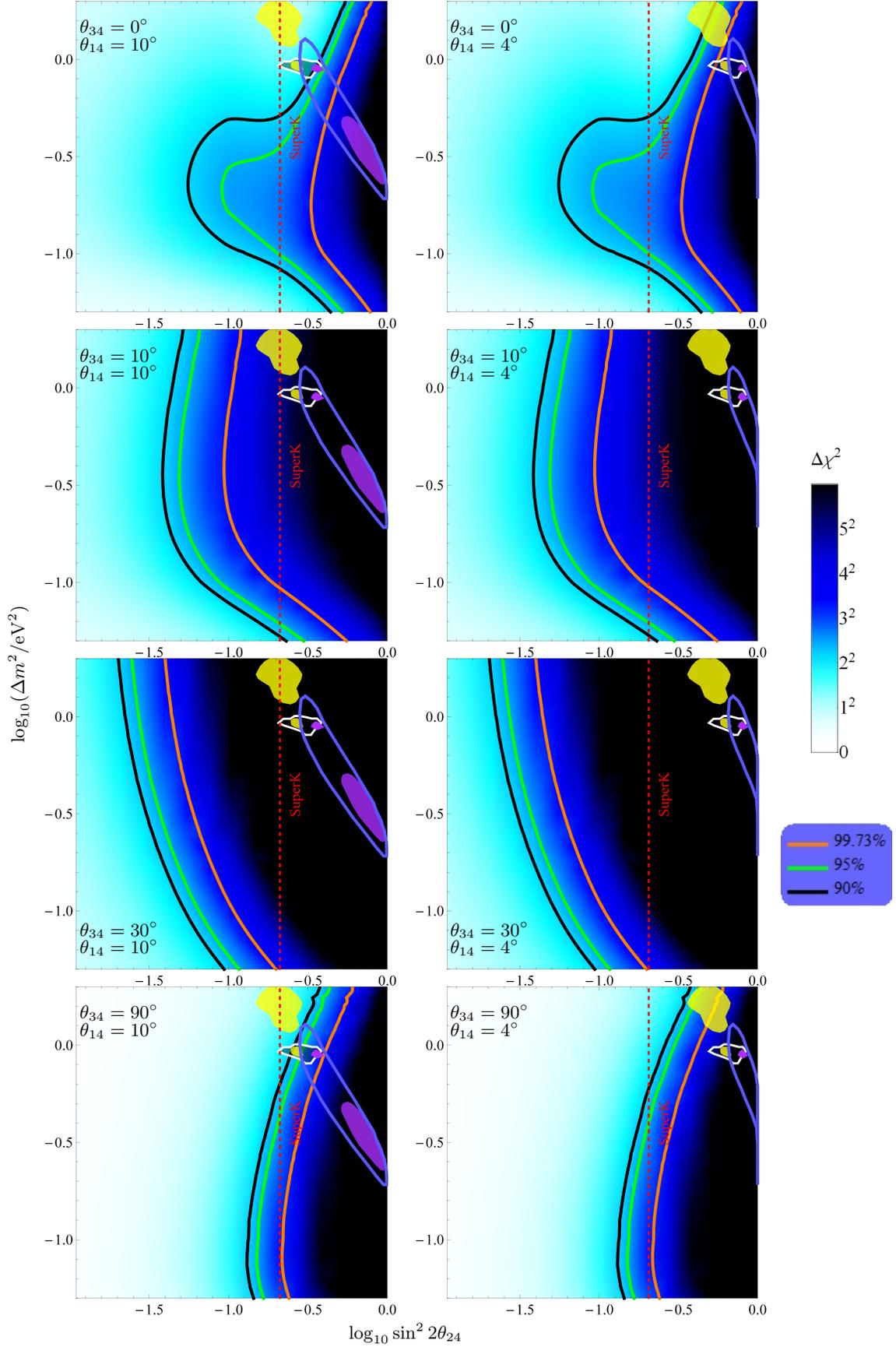


\section{$\chi^{2}$-fit and result}\label{sec:chi}

We perform now a numerical analysis of the IceCube-59 data from 
Ref.\ \cite{Aartsen:2013eka}, adopting a conventional $\chi^{2}$-function  
\begin{equation}
\chi^{2}(\theta_{14},\theta_{24},\theta_{34},\Delta m^{2},a)=\frac{a^{2}}{\sigma_{a}^{2}}+\sum_{i=c_{z}\thinspace{\rm bins}}\frac{[(1+a)N_{{\rm th},i}-N_{{\rm ob},i}]^{2}}{\sigma_{{\rm stat},i}^{2}+\sigma_{{\rm sys},i}^{2}}\,,\label{eq:0820}
\end{equation}
where $N_{{\rm ob},i}$ is the observed event number in each bin with 
statistical uncertainty $\sigma_{{\rm stat},i}^{2}$ and $N_{{\rm th},i}$ 
is the predicted event number for each bin. 
Since the events number in each bin is large enough, we can simply
take $\sigma_{{\rm stat},i}=\sqrt{N_{{\rm ob},i}}$. 
For the systematic uncertainty we take, 
according to \cite{ICRC},  $\sigma_{{\rm sys},i}=0.04N_{{\rm ob},i}$. 
The factor $1+a$ in front of $N_{{\rm th},i}$ is the normalization
factor of the full flux with a large uncertainty, for instance $\sigma_{a}=15\%$
at $10^{2}$ GeV or $25\%$ at $10^{3}$ GeV (see Fig.\ 11 in \cite{Honda:2006qj}).
The major constraint on sterile neutrinos comes from the second
term in Eq.\ (\ref{eq:0820}). For minimization of the $\chi^{2}$-function,
it is easy to compute the value of $a$ at the minimum analytically: 
\begin{equation}
a_{{\rm min}}=\frac{\sum_{i}(N_{{\rm ob},i}-N_{{\rm th},i})N_{{\rm th},i}/d_{i}}{\sigma_{a}^{-2}+\sum_{i}N_{{\rm th},i}^{2}/d_{i}}\,.\label{eq:0820-1}
\end{equation}
Here $d_{i}\equiv\sigma_{{\rm stat},i}^{2}+\sigma_{{\rm sys},i}^{2}$.
In practice, we will use Eq.\ (\ref{eq:0820-1}) for the minimization
of the $\chi^{2}$-function.

Let us first simply fix $\theta_{14}$ and $\theta_{34}$ to 
certain values so $\chi^{2}$ is a function of $\theta_{24},\thinspace\Delta m^{2}$
and $a$.  
The conventional treatment is then to replace $a$ in the $\chi^{2}$-function with 
$a_{{\rm min}}$ given by Eq.\ (\ref{eq:0820-1}). After that, $\chi^{2}$ will be a
function only of $\theta_{24}$ and $\Delta m^{2}$. 
We perform the $\chi^{2}$-fit for several cases.  
For instance, for $(\theta_{14},\theta_{34})=(10^{\circ},0^{\circ})$ 
we find the best-fit point is $\theta_{24}=8.3^{\circ}$,
$\Delta m^{2}=0.014\,{\rm eV}^{2}$ with $\chi_{{\rm min}}^{2}=9.19$. 
This is no significant effect as 
for $\theta_{24}=0^{\circ}$ (i.e., no sterile neutrinos) $\chi^{2}=9.43$ is essentially an equivalently good fit to the data. 
In Fig.\ \ref{fig:distributions} we show the event distributions of some cases where we 
can see that the best-fit curve (red, dashed) almost overlaps with the green curve (no
sterile neutrinos). \\

In Fig.\ \ref{fig:Exclusion-bound} we plot the $90\%,95\%,99.73\%$
exclusion bounds (i.e.\ contours for $\chi^{2}-\chi_{{\rm min}}^{2}=4.61,5.99,11.83$)
in the $\sin^{2}2\theta_{24}-\Delta m^{2}$ plane, for several cases. 
We also display for comparison  the 99\% C.L.\ regions of various global fit results on $\sin^2 2 \theta_{e\mu}$ from Refs.\ \cite{Kopp:2013vaa,Gariazzo:2015rra,Conrad:2012qt}. Though that signal depends on 
$\sin^{2}2\theta_{\mu e}=\sin^{2}2\theta_{14}\sin^{2}\theta_{24}$
and $\Delta m^{2}$, we can use the fixed value $\theta_{14}=10{}^{\circ}$ (or
$4^{\circ}$) to convert the LSND constraint from 
$\sin^{2}2\theta_{\mu e}$ to $\sin^{2}2\theta_{24}$.  The Super-Kamiokande result on 
$|U_{\mu 4}|^{2}$ is also given \cite{Abe:2014gda}. 

\begin{figure}[t]
\centering

\begin{overpic}[width=6cm]{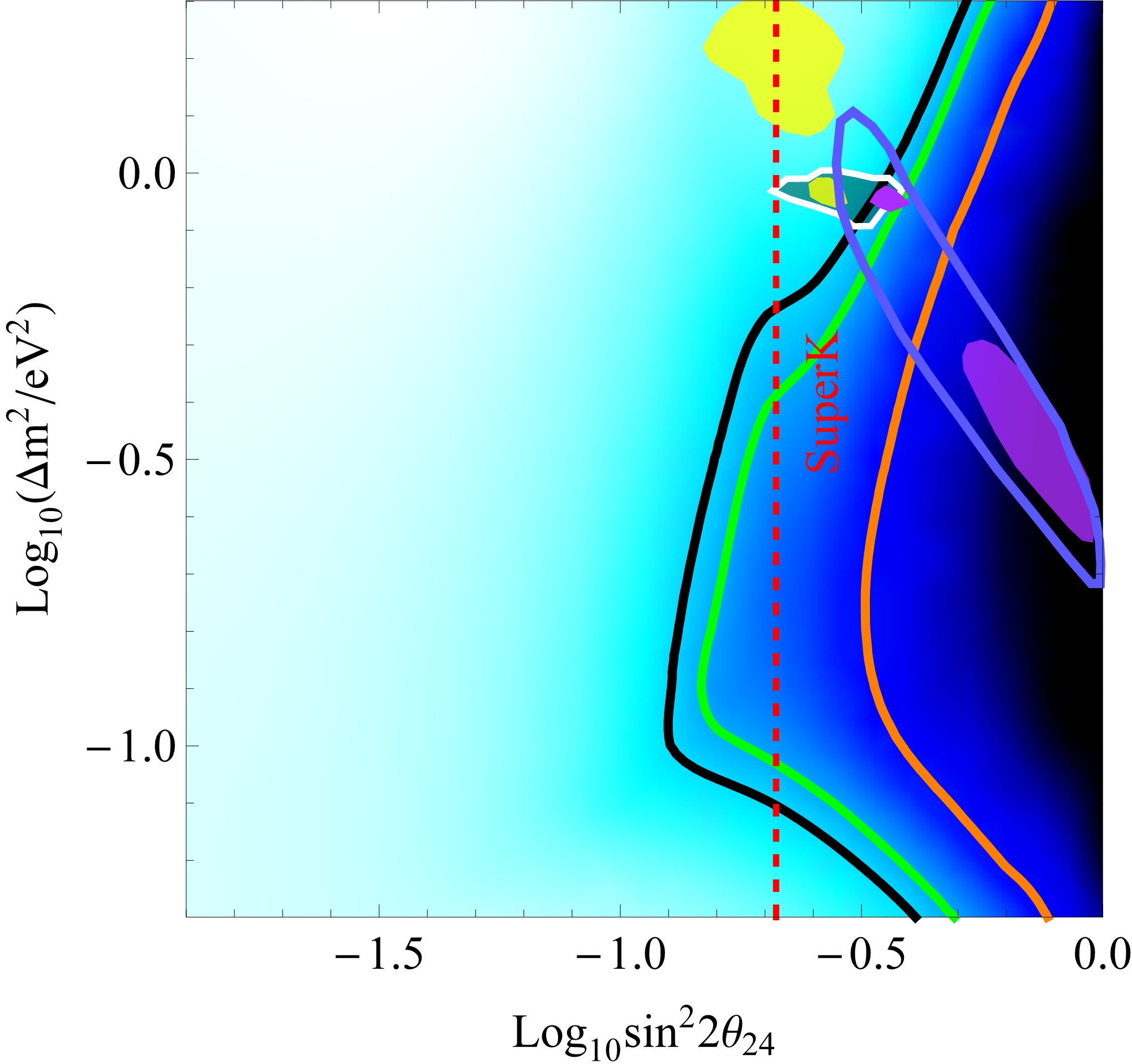}
\put (20,85) {$\theta_{14}=10^{\circ}$}
\end{overpic}
\begin{overpic}[width=6cm]{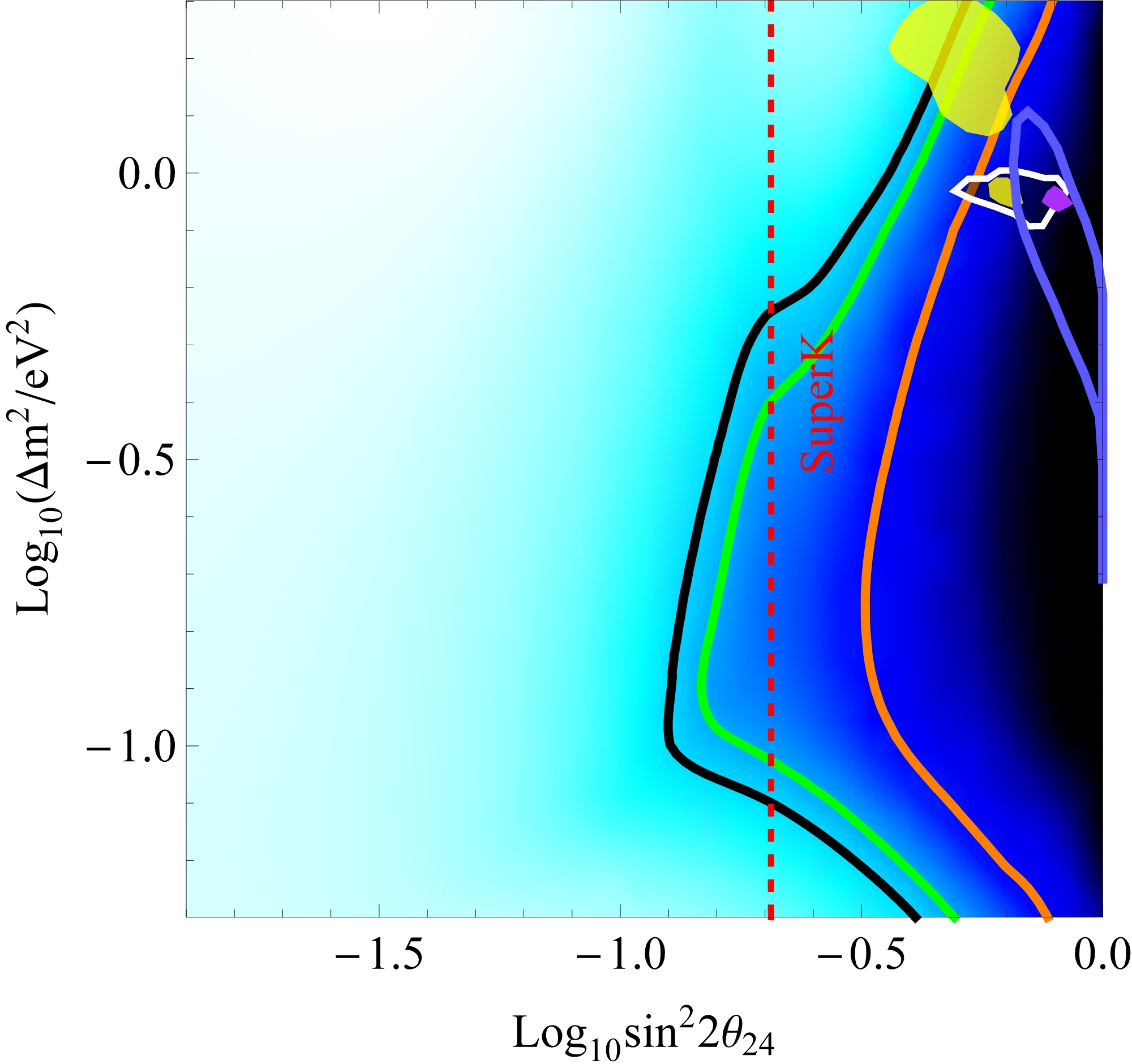}
\put (20,85) {$\theta_{14}=4^{\circ}$}
\end{overpic}
\includegraphics[height=6cm]{fig/legend}
\raisebox{2cm}{\includegraphics[width=1.5cm]{fig/sigma-legend}}

\caption{\label{fig:final}Exclusion bounds for $(\Delta m^{2},\sin^{2}2\theta_{24})$ after 
marginalizing over $\theta_{34}$; $\theta_{14}$ is fixed to $10{}^{\circ}$ (left)
or $4^{\circ}$ (right). Also shown are 99\% C.L.\ fit results from the world's appearance and disappearance 
data in purple (from Kopp \emph{et~al.\ }\cite{Kopp:2013vaa}), 
in yellow (from Giunti \emph{et~al.\ }\cite{Gariazzo:2015rra}) and in 
the white contour (from Conrad \emph{et~al.\ }\cite{Conrad:2012qt}). The blue contour line is the result of an 
appearance data only fit from \cite{Kopp:2013vaa}.}
\end{figure}

We see from Fig.\ \ref{fig:Exclusion-bound} that the value of $\theta_{34}$ is very important. 
As one increases $\theta_{34}$ from zero, the constraint becomes stronger until $\theta_{34}$ is large enough. After that, the constraint becomes weaker when $\theta_{34}$ is increased. This can be understood via Fig.\ \ref{fig:curves}, where in the bottom right panel we can see the event number generally drops down (which implies the disappearance of neutrinos) when $\theta_{34}$ is increased from $0^{\circ}$. The event number reaches its minimum at about $30^{\circ}$ and then increases because for $\theta_{34}=90^{\circ}$, as we have discussed, the signal of sterile neutrinos is weak since there is no matter effect enhancement.  
$\theta_{14}$, however, has little influence on the IceCube result for $\Delta m^{2}$ and $\sin^{2}2\theta_{24}$. But the angle is important to compare the outcome to the short baseline appearance data. 
Note that with $\theta_{14}=0$ there would be no short baseline appearance oscillations, 
see Eq.\ (\ref{eq:lsnd}). The larger $\theta_{14}$, the smaller the value of 
$\theta_{24}$ necessary to generate the same value of $\sin^2 \theta_{\mu e}$, hence it becomes more difficult for IceCube to rule it out. 
Marginalizing over $\theta_{34}$ results in 
Fig.\ \ref{fig:final}. 
We see that for $\theta_{14} = 10^\circ$ the 
appearance signal is partially compatible within $90\%$, 
but ruled out at $99.73\%$ if  
$\theta_{14} = 4^\circ$. 
Regarding the appearance plus disappearance data,  
strong dependence on the fit result exists. 
For $\theta_{14} = 10^\circ$, the signal 
from \cite{Gariazzo:2015rra} is fully consistent with IceCube-59 at $90\%$, 
whereas the full regime from \cite{Kopp:2013vaa} and a part of the regime from 
from \cite{Conrad:2012qt} are excluded at $90\%$. 
At $\theta_{14} = 4^\circ$, the signal 
from \cite{Gariazzo:2015rra} still cannot be ruled out at about $90\%$, 
whereas the full regime from \cite{Kopp:2013vaa} and the one from \cite{Conrad:2012qt} 
are ruled out at $95\%$. 
Such strong dependence on fit results and 
parameters not appearing in IceCube expressions 
will also be present in fits of IceCube-86 data, 
not yet available, and should be taken into account 
when interpreting the results.

\section{Conclusion}

We have performed a $\chi^{2}$-fit to IceCube-59 data in order to constrain sterile
neutrino mixing parameters. Special emphasis was put on the question on how muon 
neutrino disappearance compares to the muon neutrino disappearance in short baseline experiments and other sterile neutrino results. 
We have stressed the different dependence on the three relevant mixing angles and the 
important role of the very weakly constrained angle $\theta_{34}$, which governs the strength of 
the matter resonance, and is often set to zero in analyses. The value of $\theta_{14}$ is also crucial. Fixing this angle   
implies via the short baseline appearance results the value of $\theta_{24}$, 
on which the analysis of IceCube data is most 
sensitive to. It was demonstrated that only 
part of the global parameter space is currently constrained by IceCube, though of 
course important complementary information is provided. Moreover, there is dependence on which fit result one compares the data to. 

The comparison of the various oscillation channels relies on several assumptions, 
in particular the presence of only one sterile neutrino. In this framework we assume 
unitarity of the $4\times4$ mixing matrix, and the absence of additional interactions 
that sterile neutrinos could be sensitive to. As even with these assumptions 
no final solution to the short baseline disappearance and other problems can be given, it 
shows that several dedicated oscillation experiments are required 
to fully settle the issue.

\begin{acknowledgments}
We thank Christopher Wiebusch and Marius Wallraff for helpful discussions and 
Sheng-Zhi Zhao for a large contribution to the programming for 
this analysis. WR is supported by the Max Planck Society in the project
MANITOP, and by the DFG in the Heisenberg Programme with grant RO
2516/6-1; XJX is supported by the China Scholarship Council (CSC).
\end{acknowledgments}




\bibliographystyle{apsrev4-1}
\bibliography{ref}

\begin{thebibliography}{30}%
\makeatletter
\providecommand \@ifxundefined [1]{%
 \@ifx{#1\undefined}
}%
\providecommand \@ifnum [1]{%
 \ifnum #1\expandafter \@firstoftwo
 \else \expandafter \@secondoftwo
 \fi
}%
\providecommand \@ifx [1]{%
 \ifx #1\expandafter \@firstoftwo
 \else \expandafter \@secondoftwo
 \fi
}%
\providecommand \natexlab [1]{#1}%
\providecommand \enquote  [1]{``#1''}%
\providecommand \bibnamefont  [1]{#1}%
\providecommand \bibfnamefont [1]{#1}%
\providecommand \citenamefont [1]{#1}%
\providecommand \href@noop [0]{\@secondoftwo}%
\providecommand \href [0]{\begingroup \@sanitize@url \@href}%
\providecommand \@href[1]{\@@startlink{#1}\@@href}%
\providecommand \@@href[1]{\endgroup#1\@@endlink}%
\providecommand \@sanitize@url [0]{\catcode `\\12\catcode `\$12\catcode
  `\&12\catcode `\#12\catcode `\^12\catcode `\_12\catcode `\%12\relax}%
\providecommand \@@startlink[1]{}%
\providecommand \@@endlink[0]{}%
\providecommand \url  [0]{\begingroup\@sanitize@url \@url }%
\providecommand \@url [1]{\endgroup\@href {#1}{\urlprefix }}%
\providecommand \urlprefix  [0]{URL }%
\providecommand \Eprint [0]{\href }%
\providecommand \doibase [0]{http://dx.doi.org/}%
\providecommand \selectlanguage [0]{\@gobble}%
\providecommand \bibinfo  [0]{\@secondoftwo}%
\providecommand \bibfield  [0]{\@secondoftwo}%
\providecommand \translation [1]{[#1]}%
\providecommand \BibitemOpen [0]{}%
\providecommand \bibitemStop [0]{}%
\providecommand \bibitemNoStop [0]{.\EOS\space}%
\providecommand \EOS [0]{\spacefactor3000\relax}%
\providecommand \BibitemShut  [1]{\csname bibitem#1\endcsname}%
\let\auto@bib@innerbib\@empty
\bibitem [{\citenamefont {Olive}\ \emph {et~al.}(2014)\citenamefont {Olive}
  \emph {et~al.}}]{Agashe:2014kda}%
  \BibitemOpen
  \bibfield  {author} {\bibinfo {author} {\bibfnamefont {K.~A.}\ \bibnamefont
  {Olive}} \emph {et~al.} (\bibinfo {collaboration} {Particle Data Group}),\
  }\href {\doibase 10.1088/1674-1137/38/9/090001} {\bibfield  {journal}
  {\bibinfo  {journal} {Chin. Phys.}\ }\textbf {\bibinfo {volume} {C38}},\
  \bibinfo {pages} {090001} (\bibinfo {year} {2014})}\BibitemShut {NoStop}%
\bibitem [{\citenamefont {Aguilar-Arevalo}\ \emph {et~al.}(2001)\citenamefont
  {Aguilar-Arevalo} \emph {et~al.}}]{Aguilar:2001ty}%
  \BibitemOpen
  \bibfield  {author} {\bibinfo {author} {\bibfnamefont {A.}~\bibnamefont
  {Aguilar-Arevalo}} \emph {et~al.} (\bibinfo {collaboration} {LSND}),\ }\href
  {\doibase 10.1103/PhysRevD.64.112007} {\bibfield  {journal} {\bibinfo
  {journal} {Phys. Rev.}\ }\textbf {\bibinfo {volume} {D64}},\ \bibinfo {pages}
  {112007} (\bibinfo {year} {2001})},\ \Eprint
  {http://arxiv.org/abs/hep-ex/0104049} {arXiv:hep-ex/0104049 [hep-ex]}
  \BibitemShut {NoStop}%
\bibitem [{\citenamefont {Aguilar-Arevalo}\ \emph {et~al.}(2013)\citenamefont
  {Aguilar-Arevalo} \emph {et~al.}}]{Aguilar-Arevalo:2013pmq}%
  \BibitemOpen
  \bibfield  {author} {\bibinfo {author} {\bibfnamefont {A.~A.}\ \bibnamefont
  {Aguilar-Arevalo}} \emph {et~al.} (\bibinfo {collaboration} {MiniBooNE}),\
  }\href {\doibase 10.1103/PhysRevLett.110.161801} {\bibfield  {journal}
  {\bibinfo  {journal} {Phys. Rev. Lett.}\ }\textbf {\bibinfo {volume} {110}},\
  \bibinfo {pages} {161801} (\bibinfo {year} {2013})},\ \Eprint
  {http://arxiv.org/abs/1207.4809} {arXiv:1207.4809 [hep-ex]} \BibitemShut
  {NoStop}%
\bibitem [{\citenamefont {Mueller}\ \emph {et~al.}(2011)\citenamefont {Mueller}
  \emph {et~al.}}]{Mueller:2011nm}%
  \BibitemOpen
  \bibfield  {author} {\bibinfo {author} {\bibfnamefont {T.~A.}\ \bibnamefont
  {Mueller}} \emph {et~al.},\ }\href {\doibase 10.1103/PhysRevC.83.054615}
  {\bibfield  {journal} {\bibinfo  {journal} {Phys. Rev.}\ }\textbf {\bibinfo
  {volume} {C83}},\ \bibinfo {pages} {054615} (\bibinfo {year} {2011})},\
  \Eprint {http://arxiv.org/abs/1101.2663} {arXiv:1101.2663 [hep-ex]}
  \BibitemShut {NoStop}%
\bibitem [{\citenamefont {Huber}(2011)}]{Huber:2011wv}%
  \BibitemOpen
  \bibfield  {author} {\bibinfo {author} {\bibfnamefont {P.}~\bibnamefont
  {Huber}},\ }\href {\doibase 10.1103/PhysRevC.85.029901,
  10.1103/PhysRevC.84.024617} {\bibfield  {journal} {\bibinfo  {journal} {Phys.
  Rev.}\ }\textbf {\bibinfo {volume} {C84}},\ \bibinfo {pages} {024617}
  (\bibinfo {year} {2011})},\ \bibinfo {note} {[Erratum: Phys.
  Rev.C85,029901(2012)]},\ \Eprint {http://arxiv.org/abs/1106.0687}
  {arXiv:1106.0687 [hep-ph]} \BibitemShut {NoStop}%
\bibitem [{\citenamefont {Abazajian}\ \emph {et~al.}(2012)\citenamefont
  {Abazajian} \emph {et~al.}}]{Abazajian:2012ys}%
  \BibitemOpen
  \bibfield  {author} {\bibinfo {author} {\bibfnamefont {K.~N.}\ \bibnamefont
  {Abazajian}} \emph {et~al.},\ }\href@noop {} {\  (\bibinfo {year} {2012})},\
  \Eprint {http://arxiv.org/abs/1204.5379} {arXiv:1204.5379 [hep-ph]}
  \BibitemShut {NoStop}%
\bibitem [{\citenamefont {Gariazzo}\ \emph {et~al.}(2015)\citenamefont
  {Gariazzo}, \citenamefont {Giunti}, \citenamefont {Laveder}, \citenamefont
  {Li},\ and\ \citenamefont {Zavanin}}]{Gariazzo:2015rra}%
  \BibitemOpen
  \bibfield  {author} {\bibinfo {author} {\bibfnamefont {S.}~\bibnamefont
  {Gariazzo}}, \bibinfo {author} {\bibfnamefont {C.}~\bibnamefont {Giunti}},
  \bibinfo {author} {\bibfnamefont {M.}~\bibnamefont {Laveder}}, \bibinfo
  {author} {\bibfnamefont {Y.~F.}\ \bibnamefont {Li}}, \ and\ \bibinfo {author}
  {\bibfnamefont {E.~M.}\ \bibnamefont {Zavanin}},\ }\href@noop {} {\
  (\bibinfo {year} {2015})},\ \Eprint {http://arxiv.org/abs/1507.08204}
  {arXiv:1507.08204 [hep-ph]} \BibitemShut {NoStop}%
\bibitem [{\citenamefont {Aartsen}\ \emph {et~al.}(2014)\citenamefont {Aartsen}
  \emph {et~al.}}]{Aartsen:2013eka}%
  \BibitemOpen
  \bibfield  {author} {\bibinfo {author} {\bibfnamefont {M.~G.}\ \bibnamefont
  {Aartsen}} \emph {et~al.} (\bibinfo {collaboration} {IceCube}),\ }\href
  {\doibase 10.1103/PhysRevD.89.062007} {\bibfield  {journal} {\bibinfo
  {journal} {Phys. Rev.}\ }\textbf {\bibinfo {volume} {D89}},\ \bibinfo {pages}
  {062007} (\bibinfo {year} {2014})},\ \Eprint {http://arxiv.org/abs/1311.7048}
  {arXiv:1311.7048 [astro-ph.HE]} \BibitemShut {NoStop}%
\bibitem [{\citenamefont {Nunokawa}\ \emph {et~al.}(2003)\citenamefont
  {Nunokawa}, \citenamefont {Peres},\ and\ \citenamefont
  {Zukanovich~Funchal}}]{Nunokawa:2003ep}%
  \BibitemOpen
  \bibfield  {author} {\bibinfo {author} {\bibfnamefont {H.}~\bibnamefont
  {Nunokawa}}, \bibinfo {author} {\bibfnamefont {O.~L.~G.}\ \bibnamefont
  {Peres}}, \ and\ \bibinfo {author} {\bibfnamefont {R.}~\bibnamefont
  {Zukanovich~Funchal}},\ }\href {\doibase 10.1016/S0370-2693(03)00603-8}
  {\bibfield  {journal} {\bibinfo  {journal} {Phys. Lett.}\ }\textbf {\bibinfo
  {volume} {B562}},\ \bibinfo {pages} {279} (\bibinfo {year} {2003})},\ \Eprint
  {http://arxiv.org/abs/hep-ph/0302039} {arXiv:hep-ph/0302039 [hep-ph]}
  \BibitemShut {NoStop}%
\bibitem [{\citenamefont {Choubey}(2007)}]{Choubey:2007ji}%
  \BibitemOpen
  \bibfield  {author} {\bibinfo {author} {\bibfnamefont {S.}~\bibnamefont
  {Choubey}},\ }\href {\doibase 10.1088/1126-6708/2007/12/014} {\bibfield
  {journal} {\bibinfo  {journal} {JHEP}\ }\textbf {\bibinfo {volume} {12}},\
  \bibinfo {pages} {014} (\bibinfo {year} {2007})},\ \Eprint
  {http://arxiv.org/abs/0709.1937} {arXiv:0709.1937 [hep-ph]} \BibitemShut
  {NoStop}%
\bibitem [{\citenamefont {Razzaque}\ and\ \citenamefont
  {Smirnov}(2011)}]{Razzaque:2011ab}%
  \BibitemOpen
  \bibfield  {author} {\bibinfo {author} {\bibfnamefont {S.}~\bibnamefont
  {Razzaque}}\ and\ \bibinfo {author} {\bibfnamefont {A.~{\relax Yu}.}\
  \bibnamefont {Smirnov}},\ }\href {\doibase 10.1007/JHEP07(2011)084}
  {\bibfield  {journal} {\bibinfo  {journal} {JHEP}\ }\textbf {\bibinfo
  {volume} {07}},\ \bibinfo {pages} {084} (\bibinfo {year} {2011})},\ \Eprint
  {http://arxiv.org/abs/1104.1390} {arXiv:1104.1390 [hep-ph]} \BibitemShut
  {NoStop}%
\bibitem [{\citenamefont {Barger}\ \emph {et~al.}(2012)\citenamefont {Barger},
  \citenamefont {Gao},\ and\ \citenamefont {Marfatia}}]{Barger:2011rc}%
  \BibitemOpen
  \bibfield  {author} {\bibinfo {author} {\bibfnamefont {V.}~\bibnamefont
  {Barger}}, \bibinfo {author} {\bibfnamefont {Y.}~\bibnamefont {Gao}}, \ and\
  \bibinfo {author} {\bibfnamefont {D.}~\bibnamefont {Marfatia}},\ }\href
  {\doibase 10.1103/PhysRevD.85.011302} {\bibfield  {journal} {\bibinfo
  {journal} {Phys. Rev.}\ }\textbf {\bibinfo {volume} {D85}},\ \bibinfo {pages}
  {011302} (\bibinfo {year} {2012})},\ \Eprint {http://arxiv.org/abs/1109.5748}
  {arXiv:1109.5748 [hep-ph]} \BibitemShut {NoStop}%
\bibitem [{\citenamefont {Esmaili}\ \emph {et~al.}(2012)\citenamefont
  {Esmaili}, \citenamefont {Halzen},\ and\ \citenamefont
  {Peres}}]{Esmaili:2012nz}%
  \BibitemOpen
  \bibfield  {author} {\bibinfo {author} {\bibfnamefont {A.}~\bibnamefont
  {Esmaili}}, \bibinfo {author} {\bibfnamefont {F.}~\bibnamefont {Halzen}}, \
  and\ \bibinfo {author} {\bibfnamefont {O.~L.~G.}\ \bibnamefont {Peres}},\
  }\href {\doibase 10.1088/1475-7516/2012/11/041} {\bibfield  {journal}
  {\bibinfo  {journal} {JCAP}\ }\textbf {\bibinfo {volume} {1211}},\ \bibinfo
  {pages} {041} (\bibinfo {year} {2012})},\ \Eprint
  {http://arxiv.org/abs/1206.6903} {arXiv:1206.6903 [hep-ph]} \BibitemShut
  {NoStop}%
\bibitem [{\citenamefont {Esmaili}\ \emph {et~al.}(2013)\citenamefont
  {Esmaili}, \citenamefont {Halzen},\ and\ \citenamefont
  {Peres}}]{Esmaili:2013cja}%
  \BibitemOpen
  \bibfield  {author} {\bibinfo {author} {\bibfnamefont {A.}~\bibnamefont
  {Esmaili}}, \bibinfo {author} {\bibfnamefont {F.}~\bibnamefont {Halzen}}, \
  and\ \bibinfo {author} {\bibfnamefont {O.~L.~G.}\ \bibnamefont {Peres}},\
  }\href {\doibase 10.1088/1475-7516/2013/07/048} {\bibfield  {journal}
  {\bibinfo  {journal} {JCAP}\ }\textbf {\bibinfo {volume} {1307}},\ \bibinfo
  {pages} {048} (\bibinfo {year} {2013})},\ \Eprint
  {http://arxiv.org/abs/1303.3294} {arXiv:1303.3294 [hep-ph]} \BibitemShut
  {NoStop}%
\bibitem [{\citenamefont {Esmaili}\ and\ \citenamefont
  {Smirnov}(2013)}]{Esmaili:2013vza}%
  \BibitemOpen
  \bibfield  {author} {\bibinfo {author} {\bibfnamefont {A.}~\bibnamefont
  {Esmaili}}\ and\ \bibinfo {author} {\bibfnamefont {A.~{\relax Yu}.}\
  \bibnamefont {Smirnov}},\ }\href {\doibase 10.1007/JHEP12(2013)014}
  {\bibfield  {journal} {\bibinfo  {journal} {JHEP}\ }\textbf {\bibinfo
  {volume} {12}},\ \bibinfo {pages} {014} (\bibinfo {year} {2013})},\ \Eprint
  {http://arxiv.org/abs/1307.6824} {arXiv:1307.6824 [hep-ph]} \BibitemShut
  {NoStop}%
\bibitem [{\citenamefont {Arguelles}\ and\ \citenamefont
  {Kopp}(2012)}]{Arguelles:2012cf}%
  \BibitemOpen
  \bibfield  {author} {\bibinfo {author} {\bibfnamefont {C.~A.}\ \bibnamefont
  {Arguelles}}\ and\ \bibinfo {author} {\bibfnamefont {J.}~\bibnamefont
  {Kopp}},\ }\href {\doibase 10.1088/1475-7516/2012/07/016} {\bibfield
  {journal} {\bibinfo  {journal} {JCAP}\ }\textbf {\bibinfo {volume} {1207}},\
  \bibinfo {pages} {016} (\bibinfo {year} {2012})},\ \Eprint
  {http://arxiv.org/abs/1202.3431} {arXiv:1202.3431 [hep-ph]} \BibitemShut
  {NoStop}%
\bibitem [{\citenamefont {Dziewonski}\ and\ \citenamefont
  {Anderson}(1981)}]{Dziewonski:1981xy}%
  \BibitemOpen
  \bibfield  {author} {\bibinfo {author} {\bibfnamefont {A.~M.}\ \bibnamefont
  {Dziewonski}}\ and\ \bibinfo {author} {\bibfnamefont {D.~L.}\ \bibnamefont
  {Anderson}},\ }\href {\doibase 10.1016/0031-9201(81)90046-7} {\bibfield
  {journal} {\bibinfo  {journal} {Phys. Earth Planet. Interiors}\ }\textbf
  {\bibinfo {volume} {25}},\ \bibinfo {pages} {297} (\bibinfo {year}
  {1981})}\BibitemShut {NoStop}%
\bibitem [{\citenamefont {Huber}\ \emph {et~al.}(2007)\citenamefont {Huber},
  \citenamefont {Kopp}, \citenamefont {Lindner}, \citenamefont {Rolinec},\ and\
  \citenamefont {Winter}}]{Huber:2007ji}%
  \BibitemOpen
  \bibfield  {author} {\bibinfo {author} {\bibfnamefont {P.}~\bibnamefont
  {Huber}}, \bibinfo {author} {\bibfnamefont {J.}~\bibnamefont {Kopp}},
  \bibinfo {author} {\bibfnamefont {M.}~\bibnamefont {Lindner}}, \bibinfo
  {author} {\bibfnamefont {M.}~\bibnamefont {Rolinec}}, \ and\ \bibinfo
  {author} {\bibfnamefont {W.}~\bibnamefont {Winter}},\ }\href {\doibase
  10.1016/j.cpc.2007.05.004} {\bibfield  {journal} {\bibinfo  {journal}
  {Comput. Phys. Commun.}\ }\textbf {\bibinfo {volume} {177}},\ \bibinfo
  {pages} {432} (\bibinfo {year} {2007})},\ \Eprint
  {http://arxiv.org/abs/hep-ph/0701187} {arXiv:hep-ph/0701187 [hep-ph]}
  \BibitemShut {NoStop}%
\bibitem [{\citenamefont {Wallraff}\ and\ \citenamefont
  {Wiebusch}(2014)}]{Wallraff:2014qka}%
  \BibitemOpen
  \bibfield  {author} {\bibinfo {author} {\bibfnamefont {M.}~\bibnamefont
  {Wallraff}}\ and\ \bibinfo {author} {\bibfnamefont {C.}~\bibnamefont
  {Wiebusch}},\ }\href@noop {} {\  (\bibinfo {year} {2014})},\ \Eprint
  {http://arxiv.org/abs/1409.1387} {arXiv:1409.1387 [astro-ph.IM]} \BibitemShut
  {NoStop}%
\bibitem [{\citenamefont {Honda}\ \emph {et~al.}(2007)\citenamefont {Honda},
  \citenamefont {Kajita}, \citenamefont {Kasahara}, \citenamefont
  {Midorikawa},\ and\ \citenamefont {Sanuki}}]{Honda:2006qj}%
  \BibitemOpen
  \bibfield  {author} {\bibinfo {author} {\bibfnamefont {M.}~\bibnamefont
  {Honda}}, \bibinfo {author} {\bibfnamefont {T.}~\bibnamefont {Kajita}},
  \bibinfo {author} {\bibfnamefont {K.}~\bibnamefont {Kasahara}}, \bibinfo
  {author} {\bibfnamefont {S.}~\bibnamefont {Midorikawa}}, \ and\ \bibinfo
  {author} {\bibfnamefont {T.}~\bibnamefont {Sanuki}},\ }\href {\doibase
  10.1103/PhysRevD.75.043006} {\bibfield  {journal} {\bibinfo  {journal} {Phys.
  Rev.}\ }\textbf {\bibinfo {volume} {D75}},\ \bibinfo {pages} {043006}
  (\bibinfo {year} {2007})},\ \Eprint {http://arxiv.org/abs/astro-ph/0611418}
  {arXiv:astro-ph/0611418 [astro-ph]} \BibitemShut {NoStop}%
\bibitem [{\citenamefont {Fedynitch}\ \emph {et~al.}(2012)\citenamefont
  {Fedynitch}, \citenamefont {Becker~Tjus},\ and\ \citenamefont
  {Desiati}}]{Fedynitch:2012fs}%
  \BibitemOpen
  \bibfield  {author} {\bibinfo {author} {\bibfnamefont {A.}~\bibnamefont
  {Fedynitch}}, \bibinfo {author} {\bibfnamefont {J.}~\bibnamefont
  {Becker~Tjus}}, \ and\ \bibinfo {author} {\bibfnamefont {P.}~\bibnamefont
  {Desiati}},\ }\href {\doibase 10.1103/PhysRevD.86.114024} {\bibfield
  {journal} {\bibinfo  {journal} {Phys. Rev.}\ }\textbf {\bibinfo {volume}
  {D86}},\ \bibinfo {pages} {114024} (\bibinfo {year} {2012})},\ \Eprint
  {http://arxiv.org/abs/1206.6710} {arXiv:1206.6710 [astro-ph.HE]} \BibitemShut
  {NoStop}%
\bibitem [{\citenamefont {Rodejohann}\ and\ \citenamefont
  {Valle}(2011)}]{Rodejohann:2011vc}%
  \BibitemOpen
  \bibfield  {author} {\bibinfo {author} {\bibfnamefont {W.}~\bibnamefont
  {Rodejohann}}\ and\ \bibinfo {author} {\bibfnamefont {J.~W.~F.}\ \bibnamefont
  {Valle}},\ }\href {\doibase 10.1103/PhysRevD.84.073011} {\bibfield  {journal}
  {\bibinfo  {journal} {Phys. Rev.}\ }\textbf {\bibinfo {volume} {D84}},\
  \bibinfo {pages} {073011} (\bibinfo {year} {2011})},\ \Eprint
  {http://arxiv.org/abs/1108.3484} {arXiv:1108.3484 [hep-ph]} \BibitemShut
  {NoStop}%
\bibitem [{\citenamefont {Xu}(2015)}]{Xu:2015kma}%
  \BibitemOpen
  \bibfield  {author} {\bibinfo {author} {\bibfnamefont {X.-J.}\ \bibnamefont
  {Xu}},\ }\href@noop {} {\  (\bibinfo {year} {2015})},\ \Eprint
  {http://arxiv.org/abs/1502.02503} {arXiv:1502.02503 [hep-ph]} \BibitemShut
  {NoStop}%
\bibitem [{\citenamefont {Chizhov}\ \emph {et~al.}(1998)\citenamefont
  {Chizhov}, \citenamefont {Maris},\ and\ \citenamefont
  {Petcov}}]{Chizhov:1998ug}%
  \BibitemOpen
  \bibfield  {author} {\bibinfo {author} {\bibfnamefont {M.}~\bibnamefont
  {Chizhov}}, \bibinfo {author} {\bibfnamefont {M.}~\bibnamefont {Maris}}, \
  and\ \bibinfo {author} {\bibfnamefont {S.~T.}\ \bibnamefont {Petcov}},\
  }\href@noop {} {\  (\bibinfo {year} {1998})},\ \Eprint
  {http://arxiv.org/abs/hep-ph/9810501} {arXiv:hep-ph/9810501 [hep-ph]}
  \BibitemShut {NoStop}%
\bibitem [{\citenamefont {Kopp}\ \emph {et~al.}(2013)\citenamefont {Kopp},
  \citenamefont {Machado}, \citenamefont {Maltoni},\ and\ \citenamefont
  {Schwetz}}]{Kopp:2013vaa}%
  \BibitemOpen
  \bibfield  {author} {\bibinfo {author} {\bibfnamefont {J.}~\bibnamefont
  {Kopp}}, \bibinfo {author} {\bibfnamefont {P.~A.~N.}\ \bibnamefont
  {Machado}}, \bibinfo {author} {\bibfnamefont {M.}~\bibnamefont {Maltoni}}, \
  and\ \bibinfo {author} {\bibfnamefont {T.}~\bibnamefont {Schwetz}},\ }\href
  {\doibase 10.1007/JHEP05(2013)050} {\bibfield  {journal} {\bibinfo  {journal}
  {JHEP}\ }\textbf {\bibinfo {volume} {05}},\ \bibinfo {pages} {050} (\bibinfo
  {year} {2013})},\ \Eprint {http://arxiv.org/abs/1303.3011} {arXiv:1303.3011
  [hep-ph]} \BibitemShut {NoStop}%
\bibitem [{\citenamefont {Conrad}\ \emph {et~al.}(2013)\citenamefont {Conrad},
  \citenamefont {Ignarra}, \citenamefont {Karagiorgi}, \citenamefont
  {Shaevitz},\ and\ \citenamefont {Spitz}}]{Conrad:2012qt}%
  \BibitemOpen
  \bibfield  {author} {\bibinfo {author} {\bibfnamefont {J.~M.}\ \bibnamefont
  {Conrad}}, \bibinfo {author} {\bibfnamefont {C.~M.}\ \bibnamefont {Ignarra}},
  \bibinfo {author} {\bibfnamefont {G.}~\bibnamefont {Karagiorgi}}, \bibinfo
  {author} {\bibfnamefont {M.~H.}\ \bibnamefont {Shaevitz}}, \ and\ \bibinfo
  {author} {\bibfnamefont {J.}~\bibnamefont {Spitz}},\ }\href {\doibase
  10.1155/2013/163897} {\bibfield  {journal} {\bibinfo  {journal} {Adv. High
  Energy Phys.}\ }\textbf {\bibinfo {volume} {2013}},\ \bibinfo {pages}
  {163897} (\bibinfo {year} {2013})},\ \Eprint {http://arxiv.org/abs/1207.4765}
  {arXiv:1207.4765 [hep-ex]} \BibitemShut {NoStop}%
\bibitem [{\citenamefont {Abe}\ \emph {et~al.}(2015)\citenamefont {Abe} \emph
  {et~al.}}]{Abe:2014gda}%
  \BibitemOpen
  \bibfield  {author} {\bibinfo {author} {\bibfnamefont {K.}~\bibnamefont
  {Abe}} \emph {et~al.} (\bibinfo {collaboration} {Super-Kamiokande}),\ }\href
  {\doibase 10.1103/PhysRevD.91.052019} {\bibfield  {journal} {\bibinfo
  {journal} {Phys. Rev.}\ }\textbf {\bibinfo {volume} {D91}},\ \bibinfo {pages}
  {052019} (\bibinfo {year} {2015})},\ \Eprint {http://arxiv.org/abs/1410.2008}
  {arXiv:1410.2008 [hep-ex]} \BibitemShut {NoStop}%
\bibitem [{\citenamefont {Adamson}\ \emph {et~al.}(2011)\citenamefont {Adamson}
  \emph {et~al.}}]{Adamson:2011ku}%
  \BibitemOpen
  \bibfield  {author} {\bibinfo {author} {\bibfnamefont {P.}~\bibnamefont
  {Adamson}} \emph {et~al.} (\bibinfo {collaboration} {MINOS}),\ }\href
  {\doibase 10.1103/PhysRevLett.107.011802} {\bibfield  {journal} {\bibinfo
  {journal} {Phys. Rev. Lett.}\ }\textbf {\bibinfo {volume} {107}},\ \bibinfo
  {pages} {011802} (\bibinfo {year} {2011})},\ \Eprint
  {http://arxiv.org/abs/1104.3922} {arXiv:1104.3922 [hep-ex]} \BibitemShut
  {NoStop}%
\bibitem [{\citenamefont {Abbasi}\ \emph {et~al.}(2009)\citenamefont {Abbasi}
  \emph {et~al.}}]{Abbasi:2009nfa}%
  \BibitemOpen
  \bibfield  {author} {\bibinfo {author} {\bibfnamefont {R.}~\bibnamefont
  {Abbasi}} \emph {et~al.} (\bibinfo {collaboration} {IceCube}),\ }\href
  {\doibase 10.1103/PhysRevD.79.102005} {\bibfield  {journal} {\bibinfo
  {journal} {Phys. Rev.}\ }\textbf {\bibinfo {volume} {D79}},\ \bibinfo {pages}
  {102005} (\bibinfo {year} {2009})},\ \Eprint {http://arxiv.org/abs/0902.0675}
  {arXiv:0902.0675 [astro-ph.HE]} \BibitemShut {NoStop}%
\bibitem [{ICR()}]{ICRC}%
  \BibitemOpen
  \href@noop {} {}\bibinfo {note} {Talk given by M. Wallraff at ICRC 2015, The
  Hague, Netherlands, August 2015.}\BibitemShut {Stop}%
\end{thebibliography}%

\end{document}